\documentclass[a4paper, twocolumn,secnumarabic, aps, superscriptaddress,
amssymb, amsmath, nobibnotes, showpacs]{revtex4}

\usepackage{graphicx}
\usepackage{dcolumn}

\begin{document}
\title{Statistics of polymer adsorption under shear flow}

\author{Gui-Li He, Ren\'e Messina and Hartmut L\"owen \vspace{0.3cm}\\ \it\small{Institut f\"ur Theoretische Physik II, Heinrich-Heine-Universit\"at D\"usseldorf, \\Universit\"atsstrasse 1, D-40225 D\"usseldorf, Germany}}

\date{\today}

\begin{abstract}
%Using non-equilibrium Brownian dynamics computer simulations, we have investigated 
%the steady state statistics of a polymer chain adsorbed to a wall under three different shear 
%environments: i) linear shear flow in the bulk (no walls); 
%ii) shear vorticity normal to the wall (NV);
%iii) shear gradient normal to the wall (NG). 
%The probability distribution functions (PDFs) of the chain end-to-end distance and 
%its orientational angles are calculated. We find that
%the PDF of the polymer chain extension can be mapped by a simple theoretical 
%finite-extensible-nonlinear-elastic (FENE) 
%dumbbell model via two fit parameters, namely 
%an effective shear rate and an effective spring 
%constant. The tails of the angular PDFs obtained by the simulations
%are in good agreement with theoretical predictions for the bulk, 
%fitting better in a wide shear rate regime for the NV set-up and but only 
%for higher shear rates for the NG set-up.
%The latter findings for the NV and NG set-up can be qualitatively
%explained by the disturbance of the bulk fluctuation due to wall adsorption.
%Furthermore, the inner bond-bond angles
%of the chain in the different shear environments are discussed in more detail.
%Finally, the frequency of the characteristic periodic tumbling 
%motion has been investigated by simulation as well and was found to be 
%sublinear with the shear rate for the three set-ups, which extends earlier 
%results done in experiments and simulations for free and tethered polymer
%molecules without adsorption.     
Using non-equilibrium Brownian dynamics computer simulations, we have investigated
the steady state statistics of a polymer chain under three different shear
environments: i) linear shear flow in the bulk (no walls),
ii) shear vorticity normal to the adsorbing wall,
iii) shear gradient normal to the adsorbing wall.
The statistical distribution  of the chain end-to-end distance and
its orientational angles are calculated within our monomer-resolved
computer simulations.
Over a wide range of shear rates, this distribution can be mapped onto a
simple theoretical finite-extensible-nonlinear-elastic
dumbbell model with  fitted  anisotropic effective spring
constants. The tails of the angular distribution functions
are consistent with scaling predictions borrowed from the bulk dumbbell model.
Finally, the frequency of the characteristic periodic tumbling 
motion has been investigated by simulation as well and was found to be
sublinear with the shear rate for the three set-ups, which extends earlier
results done in experiments and simulations for free and tethered polymer
molecules without adsorption.
\end{abstract}


\maketitle

\section{Introduction}
Adsorption of macromolecules in flowing fluids arises in countless
practical or potential applications. For instance the process of implanting bio-tissues 
happens in the shear flow field of the blood stream and represents 
a key challenge \cite{Ratner_ARBE2004} in the cure of disabilities due to 
malfunctioning organs. In gene therapy, the transport of 
DNA to a desired target site is an another big challenge. 
\cite{Felgner_SCI_Am97_276} The control of proteins changing their biofunctionality   
during adsorption to surfaces \cite{Rusmini_BM2007} is also a quite interesting 
topic for biotechnical application. Furthermore, the influence of adsorption
under flow is important in microfluidic 
devices. \cite{Pfohl} For example, shear flow generates stretched polymer chains
which can be used for microcircuit construction from single molecules.
\cite{Braun_Nature98_391,Knobloch}  
%Most of the studies have been done only in the bulk case, 
%but polymer chain adsorption onto a substrate under shear flow 
%is much less explored \cite{Shaqfeh_JNNFM05_130_1, Ivanov_JPCB09_113_3653}.   

Experimentally, polymers under shear flow have been investigated
for many decades. \cite{Kotaka_JCP66_45, Huppler_TSR67_11,Tiu_RA95_34,Marko_MM95_28,Rivetti_JMB96_264,Smith_sci99_283,Ladoux_EPL_2000,Minko_JACS_02_124, Kirwan_NanoL04_4, Shaqfeh_JNNFM05_130_1,PNAS_Schneider2007_104,He_SoMa09} 
%and most of the work focuses on the study of chain conformations 
%\cite{Rivetti_JMB96_264, Minko_JACS_02_124, Kirwan_NanoL04_4,He_SoMa09} 
%under changing of flow environments and interactions with the substrates.   
Due to both an elongational and a rotational components 
stemming from the flow inhomogeneity,
the dynamics or statistics of sheared molecules is complex.
Tumbling motion was recently addressed experimentally in the bulk, 
\cite{Schroeder_PRL_2005,Teixeira_MM05_38,Gerashchenko_PRL_2006}
but not for adsorbed polymers.

On the theoretical side, Hinch has developed a theory for a continuous flexible string 
under shear flow, \cite{Hinch_JFM76_74} and also examined the 
effect of weak Brownian motion on the polymer. \cite{Hinch_JFM76_75}
Recently, analysing the statistics of the chain end-to-end
extension and orientation, Chertkov and co-workers have addressed the 
polymer tumbling motion under shear flow. \cite{Chertkov_JFM05_531}
Furthermore, Winkler has analytically studied the dynamics of semiflexible 
polymers under the influence of shear flow. \cite{Winkler_PRL06_97_128301}
Numerical studies based on a simple polymer {\it{dumbbell}} in 
linear shear flows have been carried out to analyze the cyclic motion
by Celani et al. \cite{Celani_EPL05_464}
and Puliafito et al. \cite{Puliafito_PD05_211}. 
As far as computer simulations on the tumbling motion are concerned, 
free chains \cite{Schroeder_PRL_2005,Aust_MM02_35_8621,Liu_P04_45_1383} 
and end-tethered polymers \cite{Rafael_PRL_2006, Mueller_EPL08_81_28002}
and confined polymers in microchannels \cite{Winkler_EPL08_83_34007}
were considered. 
  
Although the statistics of the polymer end-to-end extension and angular orientation
have been investigated experimentally and theoretically in the past, most of the relative
work is essentially focused on the bulk case. 
In this paper, our main goal is to address
these properties for an  {\it{adsorbed}} polymer chain 
under  linear shear flow. 
For a pulled adsorbed chain, a simulation study was recently performed 
in ref \cite{Serr_EPL07_78_68006}.
The problem of the adsorbed case is  more complex than in the bulk since
adsorption will interfere and compete with tumbling motion in the shear flow.
Moreover, the chain conformational behaviour will depend on the direction of the shear flow relative to the 
adsorbing substrate. For a planar adsorbing wall, there are two basic possibilities for the shear flow
direction: the shear gradient can be directed  either perpendicular or parallel to the adsorbing wall.

In this work, we consider these two cases in conjunction with the bulk system as a reference
and calculate the probability distribution function (PDF)
of the chain extension  by using monomer-resolved Brownian
dynamics computer simulations. 
We find that
the PDF of the polymer chain extension can be mapped by a simple theoretical 
finite-extensible-nonlinear-elastic (FENE) 
dumbbell model via two fit parameters, namely 
an effective shear rate and an effective spring 
constant. The tails of the angular PDFs obtained by the simulations
are in good agreement with theoretical predictions stemming from the bulk.
Surprisingly, they fit even better than in the bulk 
for a wide shear rate regime when the shear gradient is directed parallel
to the adsorbing wall.
For the situation where the shear gradient is perpendicular to the wall,
disturbances of the bulk fluctuations due to wall adsorption
lead to significant deviations from a simple dumbbell model.
We also explore the inner bond-bond angles
of the chain and the frequency of the characteristic periodic tumbling 
motion which  was found to be 
sublinear with the shear rate. This  extends earlier 
experimental and simulation studies for free and tethered polymer
molecules without adsorption. 

%A preliminary account of 
%the data was already published in \cite{He!!!!!!!!!SoftMatter 2009}    

The outline of this article is as follows: Section \ref{sec:model} describes
the details of the simulation model. Section \ref{sec:result} contains the 
results, in which section \ref{sec:extension} 
deals with the statistics of the chain extension; 
\ref{sec:angle} shows the angular degrees of freedom
(for in-adsorption-plane and off-adsorption-plane orientations, respectively); 
\ref{sec:BBangle} is discussing the shear rate dependency of the bond-bond angle; 
and the last part \ref{sec:frequency}  
is devoted to analysis of the tumbling motion frequency.
Finally, a summary of our work is presented and
discussed in section \ref{sec:conclusion}.

%%%%%%%%%%%%%%%%%%%%%%%%%%%%%%%%%%%%
\section{The model}\label{sec:model}
%%%%%%%%%%%%%%%%%%%%%%%%%%%%%%%%%%%%

%%%%%%%%%%%%%%%%%
\begin{figure}[h]
\centerline{
\includegraphics[width=1.0\columnwidth]{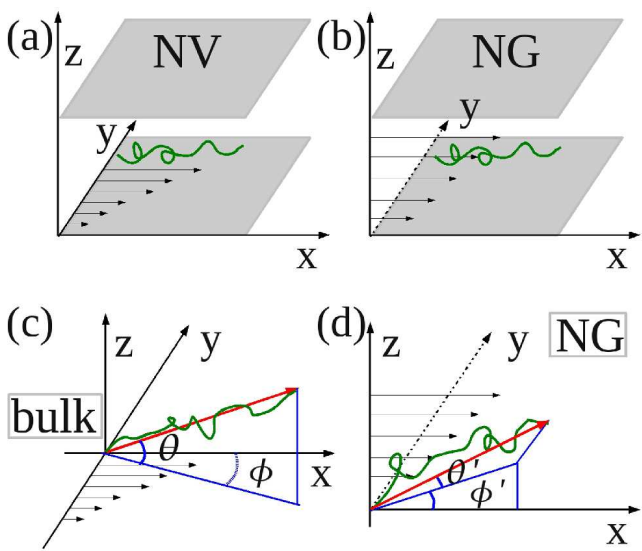}
}
\caption{Sketches of simulation model: 
(a) for the shear gradient vorticity normal to the wall (NV); 
The wall normal is along the $z-$direction.
(b) for the shear gradient normal to the wall (NG);  
(c) for the bulk (linear shear flow without walls). 
Also shown in (c) and (d) are the angles $\theta$ ($\theta'$)
and $\phi$ ($\phi'$) that characterize the end to end vector.
\label{fig:sketch}}
\end{figure}
%%%%%%%%%%%%%%%%%

This section provides details about our polymer chain model employed  
in non-equilibrium Brownian dynamics (BD) computer simulations.
\cite{Doyle_MM98_31_5474, Neelov_MTS95_4_119} 
The macromolecular chain is made up of a linear sequence of $N = 48$ 
coarse-grained bead-spring monomers, which are labeled 
$i = 0, 1, 2, \dotsm, N-1$. 
Thereby, the equation of motion of the $i$-th monomer located at a position 
${\bf r}_i(t) = (x_i(t),y_i(t),z_i(t))$ at time $t$ is 
given by the finite integration scheme  
%  
%%%%%%%%%%%%%%%  
\begin{equation}  
\label{eq:eom}  
{\bf{r}}_i(t + \delta t) = {\bf{r}}_i(t) + \frac{D_0}{k_BT}{\bf{F}}_i \delta t + \delta {\bf{G}}_i   
                           + \dot{\gamma} q_i(t) \delta t {\bf  e}_x,  
\end{equation}  
%%%%%%%%%%%%%%%  
%  
where ${\bf r}_i(t + \delta t)$ is the updated bead position at a later time $t+\delta t$. 
Hydrodynamic interactions are neglected. \cite{Hoda_JCP07_127_234902}
We now explain in details all the constitutive terms of Eq. \eqref{eq:eom}.
$D_0$, $k_B$ and $T$  stand for the free diffusion constant, 
Boltzmann's constant and the absolute temperature, respectively. 
The vector ${\bf{F}}_i$ is the total conservative force that will be fully
described later.
$\delta {\bf{G}}_i$ is the Gaussian stochastic displacement
with zero mean and variance $2D_0\delta t$ for each Cartesian component.
The last term in Eq. \ref{eq:eom} stems from the contribution of the shear flow
being in the $x$-direction, with ${\bf  e}_x$ denoting the corresponding unit 
vector, and $\dot \gamma$ the shear rate. 
$q_i(t)$ in Eq. \ref{eq:eom} can  be either $y_i(t)$ or $z_i(t)$ 
depending on the chosen shear gradient direction. 
More precisely, (i) $q_i(t)=y_i(t)$  refers to  a normal 
vorticity (NV) direction perpendicular to the adsorbing wall, as sketched in Fig.
\ref{fig:sketch}(a); 
(ii) $q_i(t)=z_i(t)$ corresponds to a normal gradient (NG), see Fig. \ref{fig:sketch}(b).
In other words, the shear gradient is in-plane with the adsorbing wall for the NV case 
whereas it becomes off-plane for the NG case.   
The aforementioned conservative force ${\bf{F}}_i$ has three contributions:
%%%%%%%%%%%%%%%%%
\begin{enumerate}
%%%%%%
\item 
Steric effects (monomer-monomer and wall-monomer repulsive interactions)
are taken into account via a truncated and shifted purely 
repulsive Lennard-Jones potential   of the form
%
%%%%%%%%%%%%%%%
\begin{equation}
\label{eq:LJ}
U_{LJ}(r) = 
4\epsilon \left[\left(\frac{\sigma_b}{r}\right)^{12}-\left(\frac{\sigma_b}{r}\right)^6
-\left(\frac{\sigma_b}{r_c}\right)^{12}+\left(\frac{\sigma_b}{r_c}\right)^6\right],   
\end{equation}
%%%%%%%%%%%%%%%
%
with a cutoff at its minimum $r_c = 2^{1/6}\sigma_b$. Here the bead diameter $\sigma_b$
and the strength of the interaction $\epsilon = k_BT$ represent the energy and length units,
respectively.

%%%%%%
\item 
The ``spring'' is modeled via a finite extensible nonlinear elastic 
(FENE)~\cite{Kremer_Grest} potential to ensure the connectivity between    
adjacent beads along the backbone, which is given by 
%
%%%%%%%%%%%%%%%%
\begin{equation}
\label{eq:fene}
U_{FENE}(r) = -\frac{1}{2} KR^2_0\ln\left[1-\left(\frac{r}{R_0}\right)^2\right] \;,
\end{equation}
%%%%%%%%%%%%%%%%
%
where the FENE cut-off (i.e. the allowed maximum bond length) is set to 
$R_0 = 1.5\sigma_b$ and the spring constant is $K = 27\epsilon/{\sigma_b}^2$. 

%%%%%%
\item 
To mimic the adsorption process, 
a strongly attractive wall defining the $(x,y)$-plane at $z = 0$ is considered.   
Thereby, we employed a Van der Waals-like attractive potential 
%
%%%%%%%%%%%%%%%%
\begin{equation}
\label{eq:ad_wall}
U_{ads}(z) = - A_0 \epsilon\left(\frac{\sigma_b}{z}\right)^6
\end{equation}
%%%%%%%%%%%%%%%%
%  
with $A_0 = 5$. A more detailed modeling for a realistic surface
was used e.g.in ref \cite{Andrienko_MM05_38_5810}.
%Here surface corrugation is neglected which is a good approximation  
%for the cleaved mica surfaces in the experiment \cite{He_SoMa09}.

\end{enumerate}
%%%%%%%%%%%%%%%

\begin{figure}[b]
\centerline{
\includegraphics[width=1.0\columnwidth]{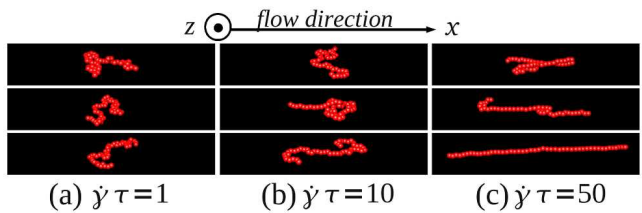}}
\caption{Typical snapshots of polymer chains under linear shear flow
in $x-$direction. The shear gradient direction is normal to the 
wall-plane (NG). Here the three columns (a)(b)(c) show configurations
for three shear rates $\dot \gamma \tau = 1$, $10$, $50$.
\label{fig:snapshots}}
\end{figure}

The BD time step is set to $\delta t = 2\times10^{-6}\tau$, 
where $\tau = \sigma_b^2/D_0$ sets the time unit.   
Typically $5 \times 10^8$ BD time steps are used for data production,
during which the first $5 \times 10^7$ 
steps are for the chain relaxation. $^{[{\bf{Note}}\ 1]}$
\footnotetext[1]{
Given the strong adsorption,
the NG case (normal shear gradient) requires more computational time 
for good statistics, so that even $10^9$ BD steps were used.}
%
%Simulation snapshots and the relative experimental AFM chain images 
%are obtained in an earlier publication \cite{He_SoMa09}. 
%Here we should also note that in the following figures
%(in section \ref{sec:result}), the error bars were calculated over five
%(for the bulk case) and ten (for the NV, NG) independent simulation runs.      
% 
Simulation snapshots are viewed in Fig. \ref{fig:snapshots}.

%%%%%%%%%%%%%%%%%%%%%
\section{Results}
\label{sec:result}
%%%%%%%%%%%%%%%%%%%%%

%%%%%%%%%%%%%%%%%%%%%%%%%%%%%%%%%%%%%%%%%%%%%%%%%%%%%%
\subsection{Statistics of the polymer extension}
\label{sec:extension}
%%%%%%%%%%%%%%%%%%%%%%%%%%%%%%%%%%%%%%%%%%%%%%%%%%%%%%

We consider a single polymer chain which is advected by a shear flow and stretched
by velocity inhomogeneity. The degree of the polymer stretching is
characterized by the chain's end-to-end distance $R_e$, for which we have
calculated the probability distribution function (PDF) under
different shear flow geometries.

\begin{figure}[t]
\centerline{
\includegraphics[angle=270,width=0.49\columnwidth]{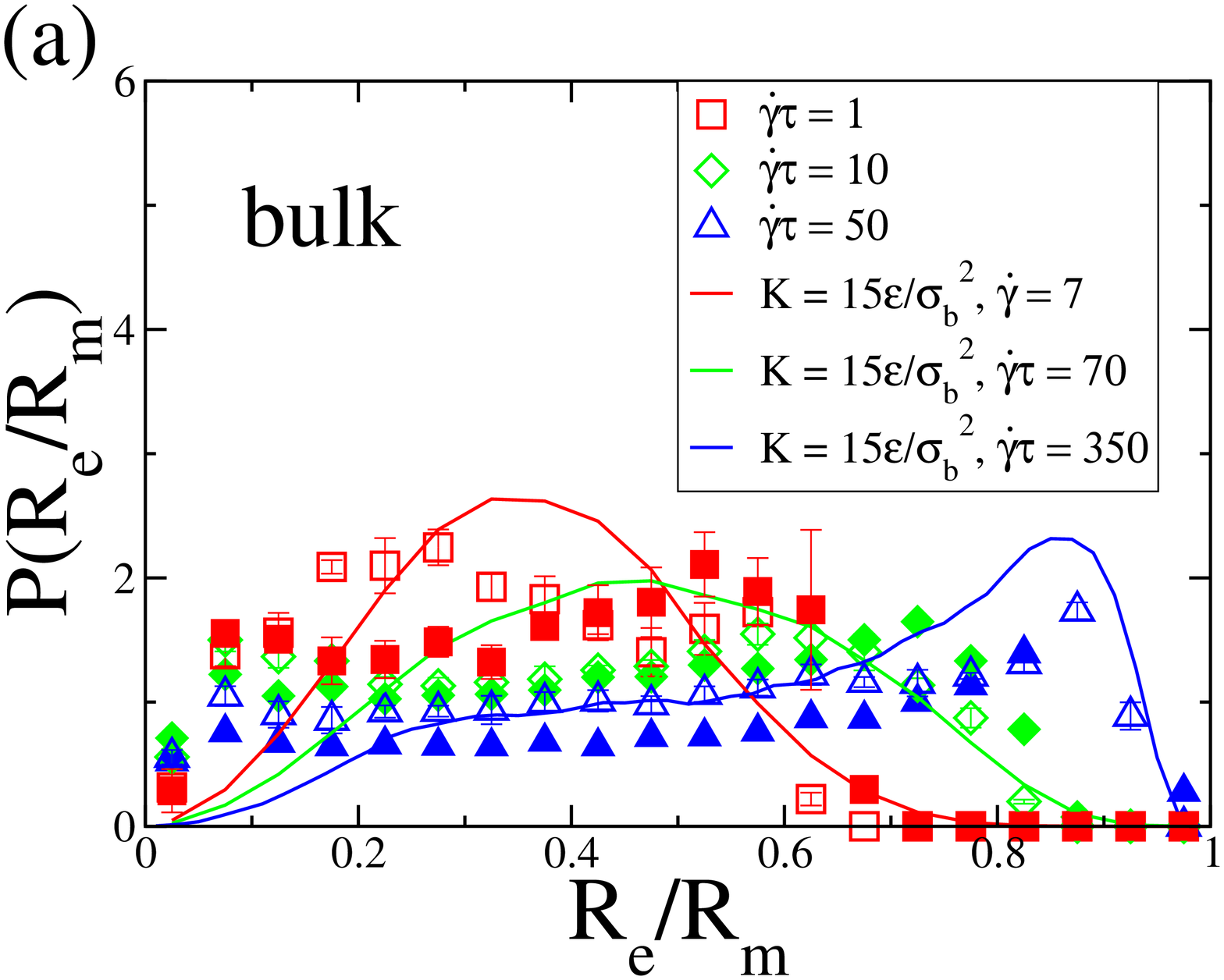}
\includegraphics[angle=270,width=0.49\columnwidth]{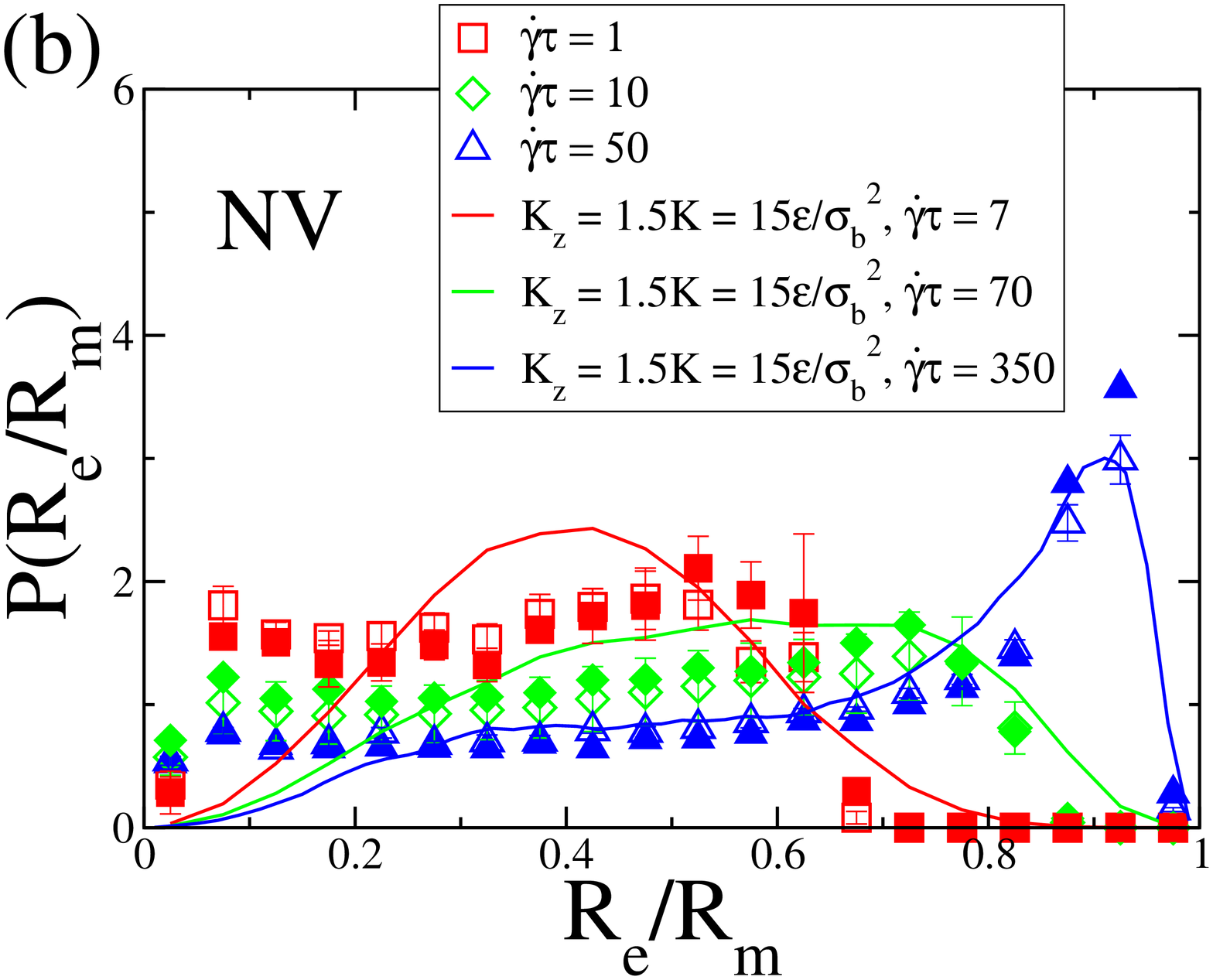}
}
%\centerline{
%\includegraphics[angle=270,width=0.4\columnwidth]{P_Re_y_fenefit.ps}
%}
\centerline{
\includegraphics[angle=270,width=0.5\columnwidth]{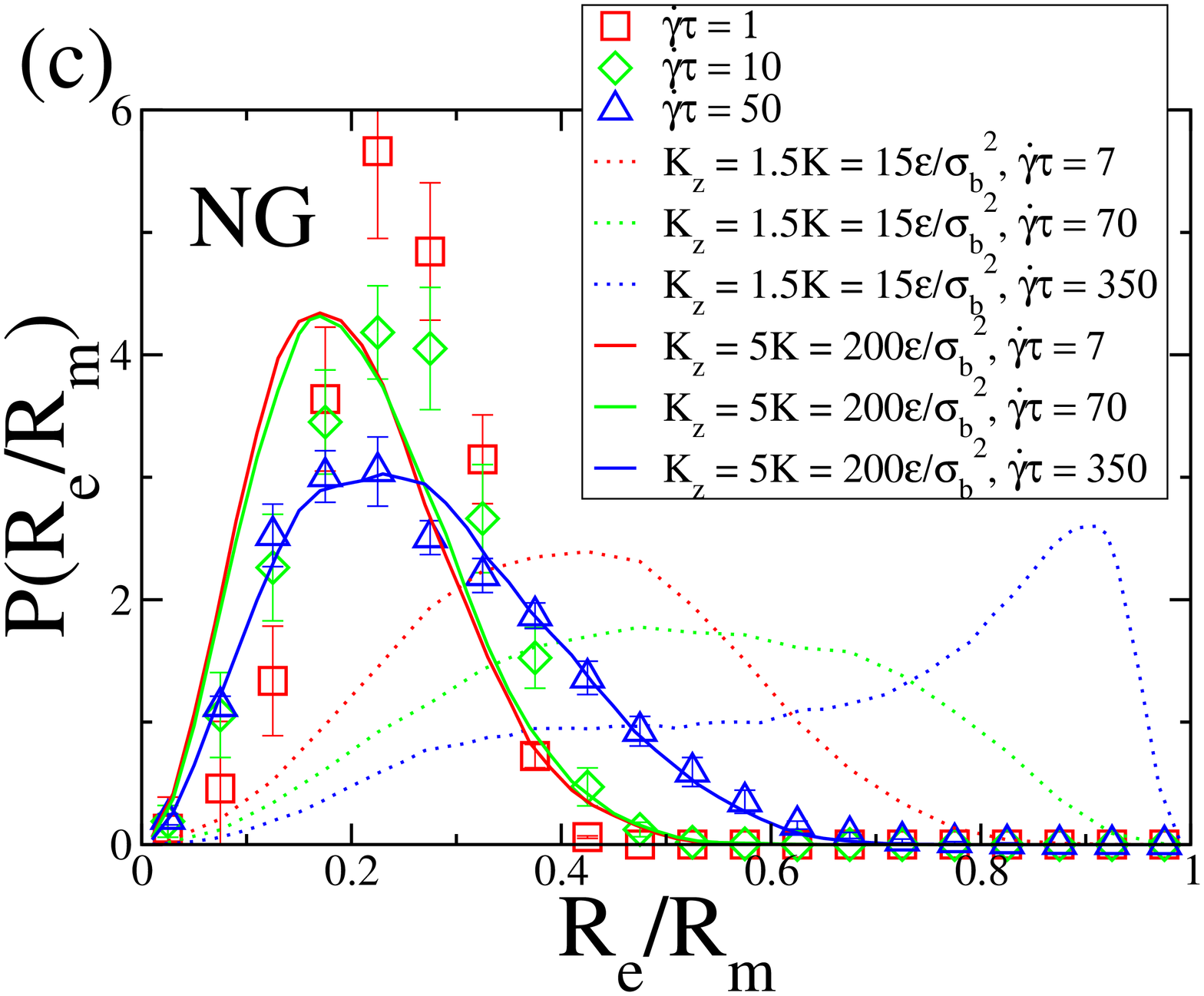}
}
\caption{The probability distribution functions  $P(R_e/R_m)$ 
of the reduced end-to-end distance $R_e/R_m$ 
for (a) the bulk, (b) the normal vorticity (NV) case and 
(c) the normal gradient (NG). 
The symbols $\square$, $\diamond$, $\triangle$ stand for the shear rate 
$\dot \gamma \tau = 1, 10, 50$, respectively,
where in (a) and (b), the pure two-dimensional (2D) case is shown as a reference  (filled symbols). 
For the FENE-dumbbell model (curves), the chosen fitting spring constant(s) parameter(s)
is(are) indicated. 
\label{fig:pdfRe}}
\end{figure}

The normalized PDF of the rescaled size extension $R_e/R_m$, 
where $R_m = (N - 1)R_0$ is the fully stretched
chain length, is plotted in Fig. \ref{fig:pdfRe} for three typical selected shear rates 
($\dot \gamma \tau = 1 (\square), 10 (\diamond), 50 (\triangle)$) for three geometries: 
(a) bulk, (b) NV, in-plane shear gradient and (c) NG, vertical shear gradient.
$^{[{\bf{Note}}\ 2]}$ 
\footnotetext[2]{In the figures of the results section, all the error bars 
were calculated using five (for the bulk case) and ten (for  NV and NG) 
independent simulation runs.}

A general feature is that upon increasing the shear rate, 
chain stretching is always favored. 
In terms of size distribution, this shear-rate induced stretching 
manifests itself either as a shift of the peak to larger sizes (see Fig. \ref{fig:pdfRe}(a,b)),
and/or as a thickening of the tail (see Fig.\ref{fig:pdfRe}(c)).    
The strongest stretching, at prescribed reduced shear rate $\dot \gamma \tau$, 
is obtained for the NV case. 
This feature can be simply explained as follows.
The NV case, due to the rather strong adsorption, leads to a chain swelling
(in the lateral dimensions) compared to the bulk conformations already at equilibrium 
(i.e., without shear flow).  
To elucidate this idea, we have also systematically shown
the {\it purely two-dimensional} case (filled symbols) in Fig. \ref{fig:pdfRe}(a,b)
as a reference. 
Moreover, for the NV case,  
the shear flow deforms the polymer chain much more  efficiently  
than in the NG case, where the shear rate gradient direction is off plane.
Clearly, deformations occur in the NG case because of temperature-induced fluctuations
in the heights of the monomers. 
  
%The obtained results from the simulations
%seemly have no qualitative difference of the PDF shape changing in the wider regime
%of shear rates between the bulk case the the NV one,
%but the chain is more stretched in the NV case than in the bulk when it is exposed to
%the same shear rate. To explain this, we have also simulated the pure  
%two-dimensional (2D) bulk model and plotted it in Fig. \ref{fig:pdfRe}(a)(b) as
%a reference. Indeed, the NV (Fig. \ref{fig:pdfRe}(b), open
%symbols) case is much closer to the 2D bulk (Fig. \ref{fig:pdfRe}(b), solid
%symbols), which implies that the NV is nearly a 2D system. This implies
%that the adsorption reduces the monomer mobility in the $z-$direction and
%the fluctuations in the shear gradient $y-$direction are increased,
%such that the chain is more strongly stretched in this case (NV) comparing
%to that in the bulk. However the PDF of the NG case (see Fig. \ref{fig:pdfRe}(c)) 
%deviates from both the bulk (Fig. \ref{fig:pdfRe}(a)) and the NV (Fig. \ref{fig:pdfRe}(b))
%cases, where the peaks are in the left side which means more coil-sized
%conformations appearing in the NG model. Since the strong adsorption induced from
%the bottom wall, the monomers become less sensitive to the vertical shear
%gradient flow which causes the chains are less stretched by this velocity
%inhomogeneity.

%%%%%%%%%%%%%%%%%%
%%%%%%%%%%%%%%%%%%

To broaden our understanding of the statistics of the polymer 
chain under shear flow, we have additionally considered the well
known so-called FENE-dumbbell model. \cite{Celani_EPL05_464} 
The equation, describing the evolution of the end-to-end vector ${\bf{R}}_e$ 
of a FENE-dumbbell (i.e., a dimer), is given by
%
%%%%%%%%%%%%%%%%
\begin{equation}
\label{eq:feneRe}
{\bf{R}}_e(t + \delta t) = {\bf{R}}_e(t) + \frac{D_0}{k_BT}{\bf{F}}_{FENE} \delta t + \delta {\bf{G}}   
                           + \dot{\gamma} Q(t) \delta t {\bf  e}_x,
\end{equation}
%%%%%%%%%%%%%%%%
%
where ${\bf{F}}_{FENE} = - \nabla U_{FENE}(R_e)$ is the FENE force;
$\delta {\bf{G}}$ is the Gaussian term as in eq. \ref{eq:eom};
$Q$ is the $y-$ (or $z-$) component $R_{ey}$ (or $R_{ez}$) of ${\bf{R}}_e$ 
assuming the shear gradient applied in the $y-$ (or $z-$) direction.
To take adsorption into account in the most simple and physically sound manner within the dumbbell model, 
we have introduced anisotropic FENE-spring constants such that
$K_x = K_y = K_z = K$ in the bulk and $[K_z > (K_x = K_y = K)]$
for NV and NG cases. 
The stochastic equation \eqref{eq:feneRe} was solved by using a 
straightforward one-particle BD scheme integration.
One has to bear in mind that this dumbbell approach is certainly relevant in the limit 
of strong stretching, where the degrees of freedom of ``inner'' monomers of a real polymer 
chain would be indeed heavily reduced, 
but can not be suitable at low shear rates where
the real chain size is dictated by the complicated distribution of the monomers.

By choosing a suitable choice of parameters (effective spring constant(s), effective shear rate) 
for the dumbbell model,  
one could always perfectly match any PDF of the fully monomer resolved chain (results not shown).
However, in order to make the comparison more physical
between the dumbbell model and the fully monomer resolved chain,
we kept the spring dumbbell constant fixed as well as the ratio between the shear rates,
see Fig. \ref{fig:pdfRe} ($7 : 70 : 350 = 1 : 10 : 50$).
Given the very crude representation of the chain via a dumbbell,
Fig. \ref{fig:pdfRe}(a,b) shows a remarkable good agreement between 
BD data and the dumbbell model, especially at high shear rates and/or large chain size.
The good agreement between these two approaches in Fig. \ref{fig:pdfRe}(b) for the NV case 
also demonstrates that the idea of mimicking the adsorption with a larger dumbbell-spring constant 
in the $z$-direction is fruitful. 
Nonetheless, the NG case is too complex to be captured by our simple dumbbell model.
Indeed, physically, the deformation in this situation is due to the $z$-fluctuations of 
the chain-monomers which can not be suitably approximated by a dumbbell.
However, if we consider a stiffer (effective) dumbbell  with 
$K_z=5K=200$ for the NG case, we can recover again a good agreement with the
BD data, see Fig. \ref{fig:pdfRe}(c).

%
%%%%%%%%%%%%%%%%% 
\begin{figure}[t]
\centerline{
\includegraphics[angle=270,width=0.58\columnwidth]{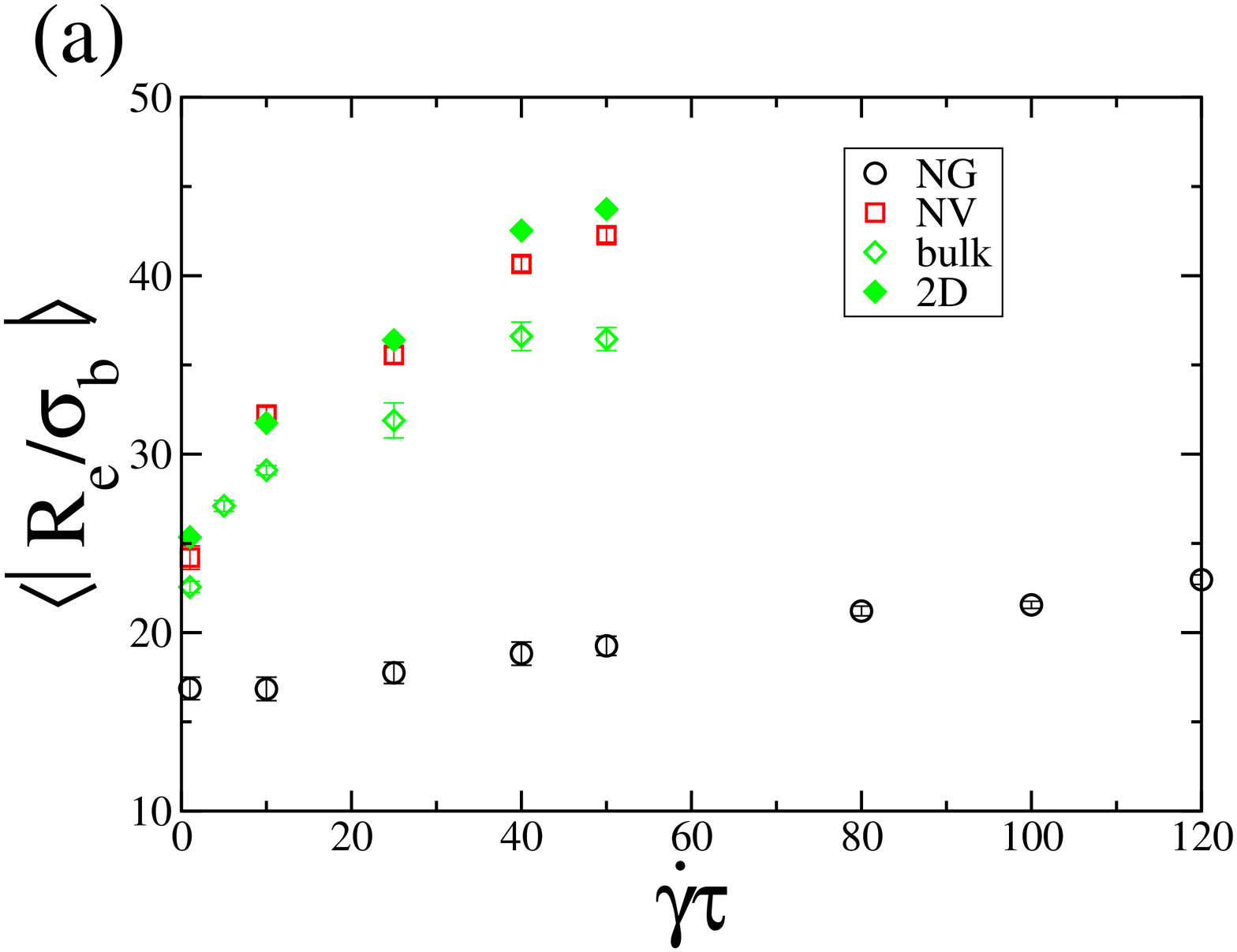}
\includegraphics[angle=270,width=0.58\columnwidth]{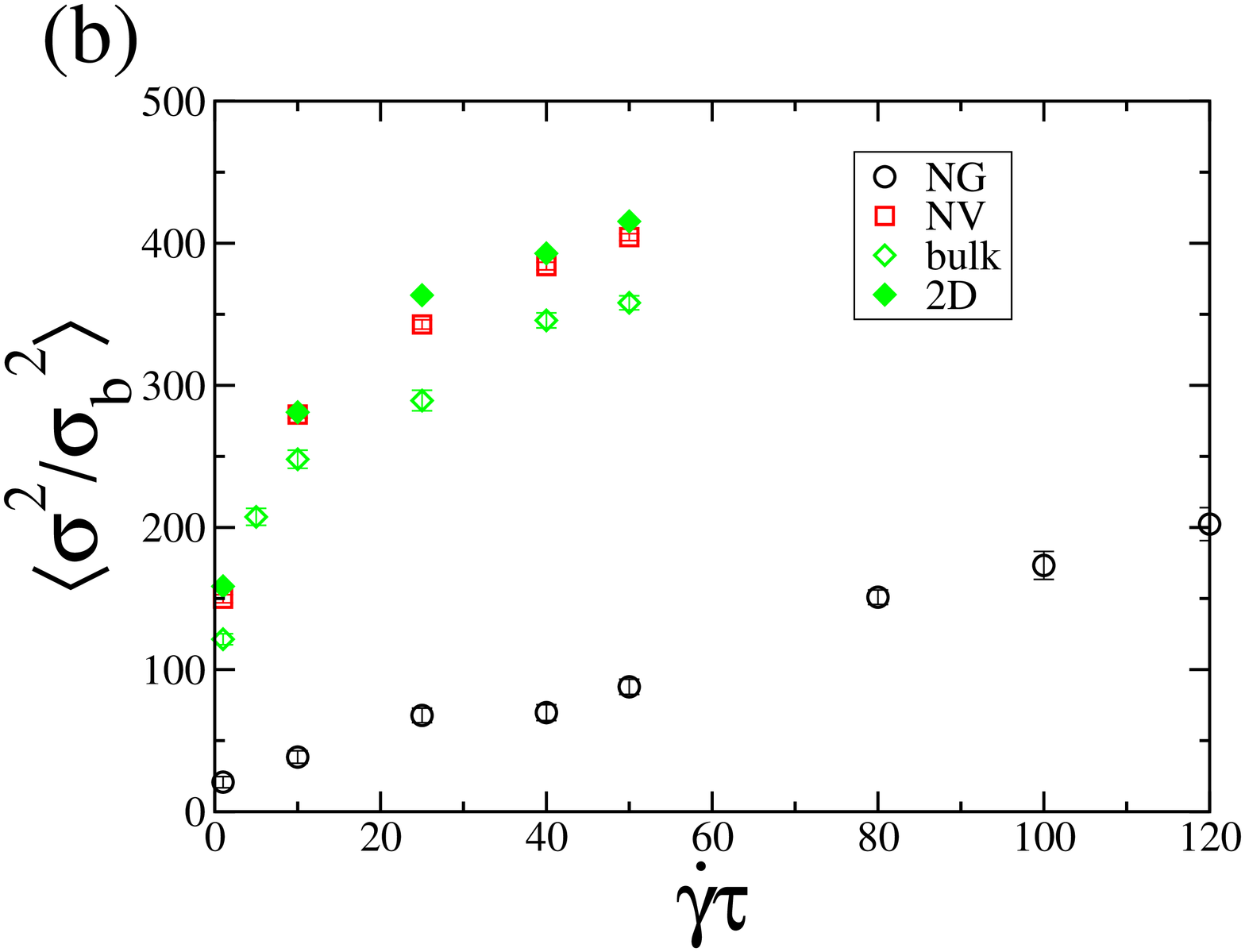}
}
%\centerline{
%\includegraphics[angle=270,width=0.87\columnwidth]{second_Re_3cases_err.ps}
%}
\centerline{
\includegraphics[angle=270,width=0.6\columnwidth]{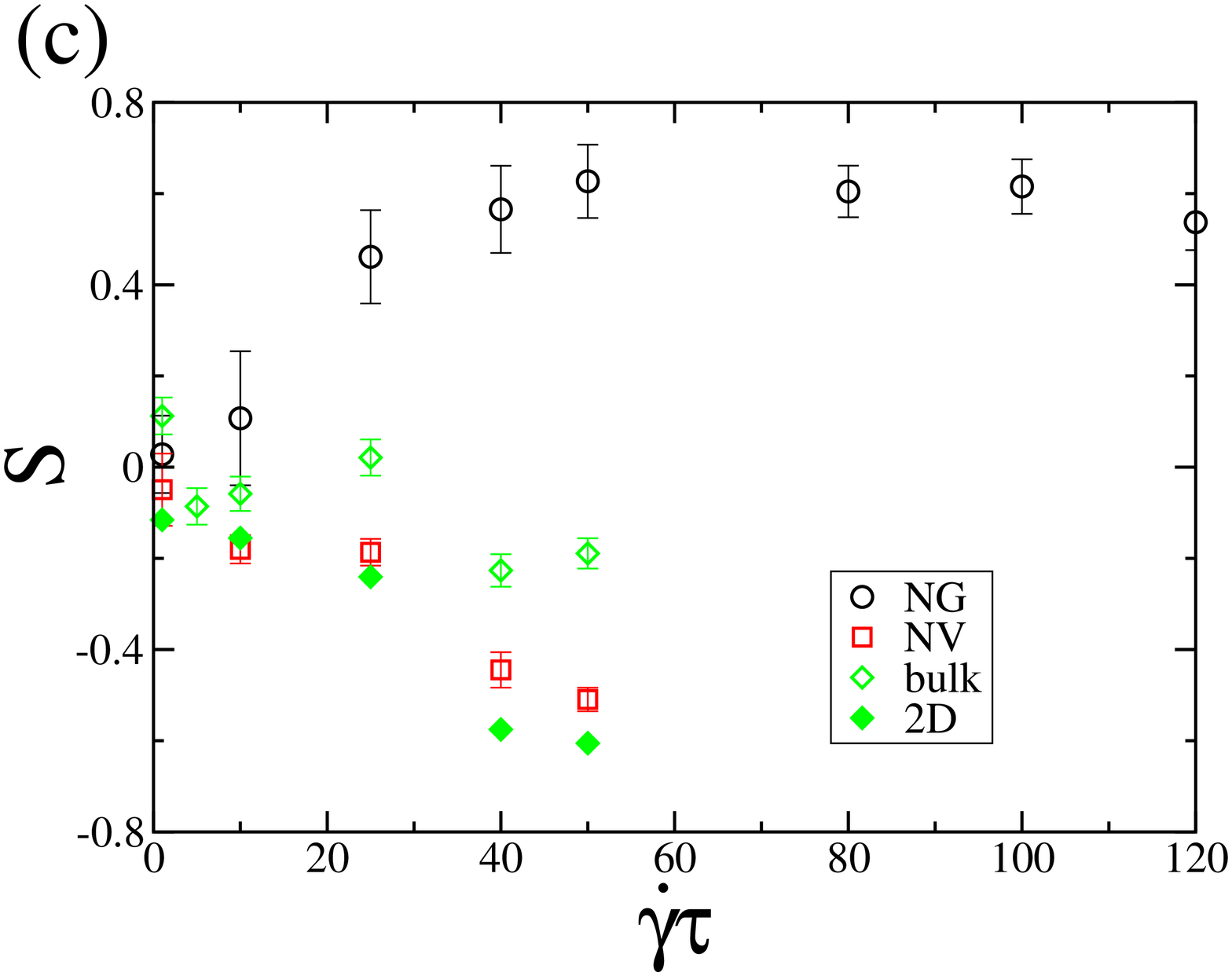}
}
\caption{The first three moments of the probability distribution function
of end-to-end distance (see Fig. \ref{fig:pdfRe}) for the
three cases (NG, NV and bulk) plotted against the shear rate: 
(a) the mean value of the end-to-end distance $\left<|R_e|\right>$;
(b) the variance $\left<\sigma^2\right>$;
(c) the third standardized moment ($S$), so-called skewness. 
Here the solid symbols stand for the purely 2D results.  
\label{fig:momentsRe}}
\end{figure} 
%%%%%%%%%%%%%%%%% 
      
To further characterize the statistics of the adsorbed chain under shear flow, 
we have calculated the first three moments of the end-to-end size distributions 
(the PDF in Figure \ref{fig:pdfRe}), which are 
(i) the mean value $\left<|{\bf{R}}_e|\right> = \left<R_e\right>$,
(ii) the variance $\sigma^2$, and
(iii) the normalized third moment (so-called {\it{skewness}}, $S$). 
%and plotted inFig. \ref{fig:momentsRe}(a)(b)(c) correspondingly.   

The mean value of the end-to-end distance is obtained as follows
%
%%%%%%%%%%%%%%%%
\begin{equation}
\label{mean}
\left<|{\bf{R}}_e|\right> = \left<R_e\right>  = \left<\sum_{m = 1}^MR_{em}\right>
 =  \frac{\sum_{m = 1}^M|{\bf{r}}_{N - 1} - {\bf{r}}_0|_m}{M} ,
\end{equation}
%%%%%%%%%%%%%%%%
%
where $R_{em}$ is the absolute value of the end-to-end vector in the 
$m-$th configuration and the sum runs over the total number of the configurations $M$.
%of $M = 9200$ for $\dot \gamma \tau \leq 25$ in the NG case
%as well as $M = 4200$ for the rest. 
Figure \ref{fig:momentsRe}(a) shows that $\left<R_e\right>$ grows  
with increasing  ($\dot \gamma \tau$), as expected.
Nevertheless, in the NG case, the shear rate dependency
of the end to end distance is relatively weak compared to the other
flow geometries.
  
The second moment, which describes the degree of broadening
about the mean value, is defined as follows
%
%%%%%%%%%%%%%%%%
\begin{equation}
\label{variance}
\sigma^2 = \frac{\sum_{m = 1}^M\left(R_{em} - \left<R_e\right>\right)^2}{M} \;.
\end{equation}
%%%%%%%%%%%%%%%%
%
Figure \ref{fig:momentsRe}(b) clearly shows that the variance 
$\sigma^2$ grows with increasing shear rate. 
This effect is actually due to the tumbling (more on that later) allowing the chain 
to cyclically stretch and recover its coil-like conformation.
Thereby, the higher the shear rate  the more stretched the chain can be,
and consequently the larger the variance becomes.
In the NG case, where stretching at prescribed shear rate is much weaker,
the broadening of the size distribution is reduced accordingly.

The skewness $S$, a central quantity in probability theory and statistics, \cite{skewness} 
is  defined as $S = \mu_3/\sigma^3$,  where $\mu$ is the third moment with respect
to the mean, and $\sigma$ is the standard deviation. 
Explicitly, the skewness was computed as follows
%
%%%%%%%%%%%%%%%%
\begin{equation}
\label{3id_skew}
S = \frac{\mu_3}{\sigma^3} = \frac{\frac{1}{M}\sum_{m = 1}^M\left(R_{em} - \left<R_e\right>\right)^3} 
{\left[\frac{1}{M}\sum_{m = 1}^M\left(R_{em} - \left<R_e\right>\right)^2\right]^{3/2}} \;,
\end{equation}
%%%%%%%%%%%%%%%%
%
It provides a quantitative and simple measure of the asymmetry of 
the probability distribution function.
Hence, the qualitative difference between the PDF in the NG model (see Fig. \ref{fig:pdfRe}(c)) 
and those found in the other cases (see Fig. \ref{fig:pdfRe}(a)(b)) can be rationalized
in terms of the value as well as the sign of $S$, see Fig. \ref{fig:momentsRe}(c). 
For the shear gradient normal to the wall (NG),  Fig. \ref{fig:momentsRe}(c) shows
that $S$ is positive and increases with shear rate. The positivity of ($S$) is a
signature of the statistical preponderance of the coil-sized chain configurations in the NG model. 
On the other hand, the skewness becomes 
negative and decreases (i.e., its magnitude increases) with growing 
shear rate in the NV model.
This feature clearly indicates the preponderance of stretched chain conformations. 
%
%The physical origin of the sign difference in $S$ for
%the PDF of the end-to-end distance at different shear gradient directions (NG, NV) 
%is the same as that controlling the polymer size distribution, which was explained
%in our earlier work \cite{He_SoMa09}. The skewness of the chain size distribution
%in the three-dimensional (3D) and two-dimensional bulk under linear shear flow
%is plotted as well in Fig. \ref{fig:momentsRe}(c) (open diamonds for the 3D and the
%solid diamonds for the 2D, correspondingly). Clearly,
%the 3D bulk case is different from the confined ones.

%%%%%%%%%%%%%%%%%%%%%%%%%%%%%%%%%%%%%%%%%%%%%%%%%%%%%%%%%%%%%%%%%%%%%%%%%%%%%%%%    
\subsection{Statistics of the polymer end-to-end vector orientations}
\label{sec:angle}
%%%%%%%%%%%%%%%%%%%%%%%%%%%%%%%%%%%%%%%%%%%%%%%%%%%%%%%%%%%%%%%%%%%%%%%%%%%%%%%%    

%%%%%%%%%%%%%%%%%
\begin{figure}[h]
\centerline{
\includegraphics[angle=270,width=0.46\columnwidth]{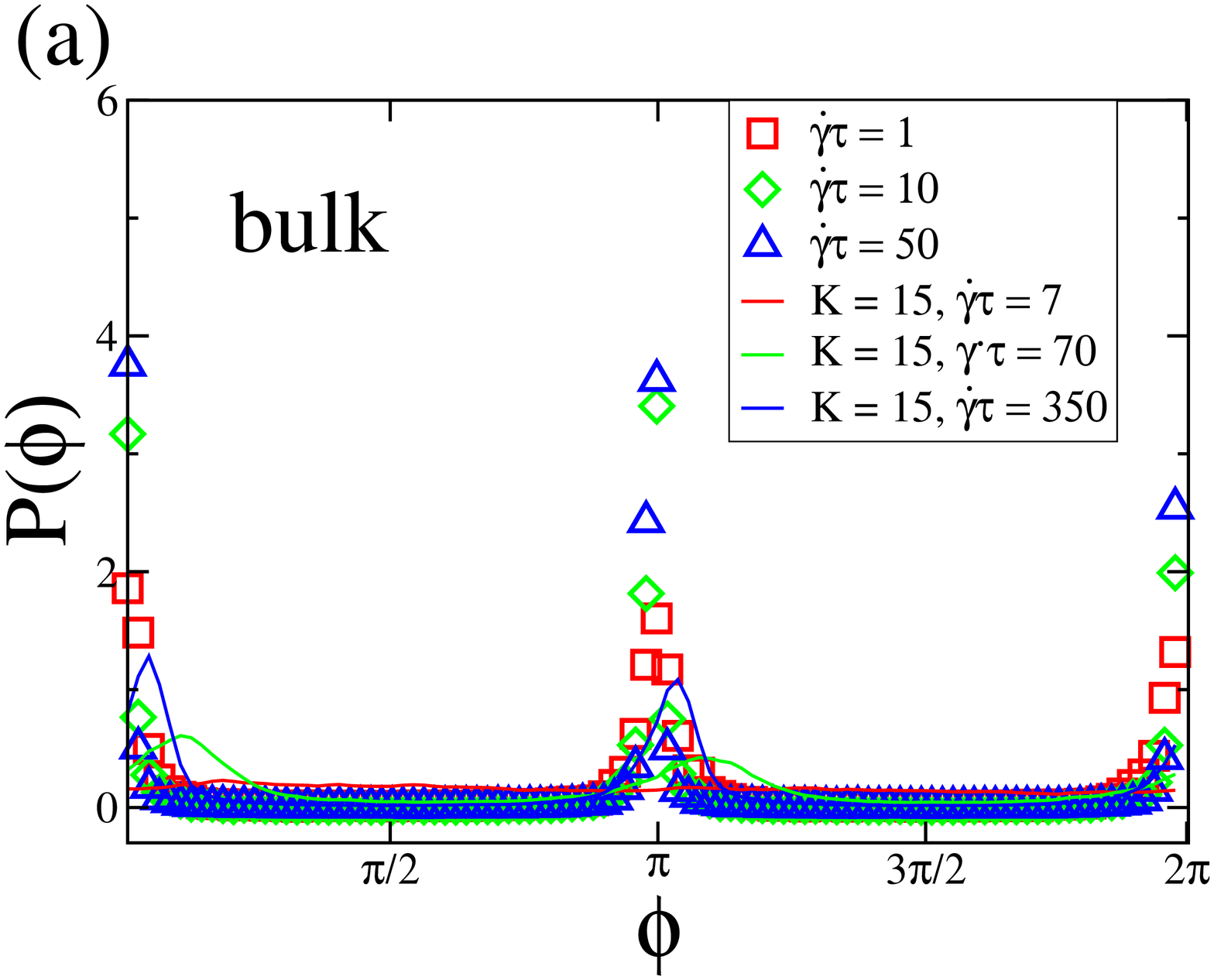}
\includegraphics[angle=270,width=0.46\columnwidth]{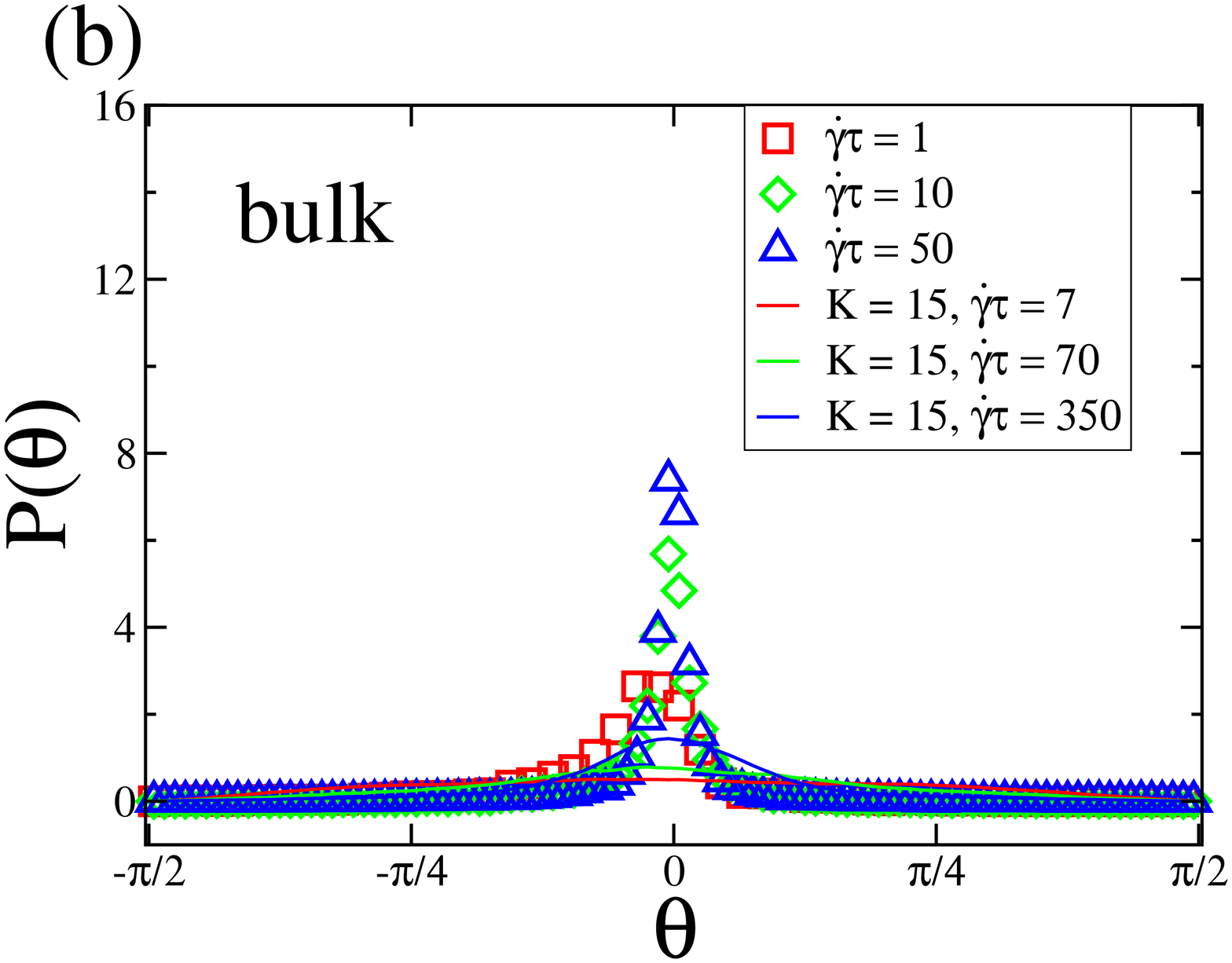}
}
\centerline{
\includegraphics[angle=270,width=0.46\columnwidth]{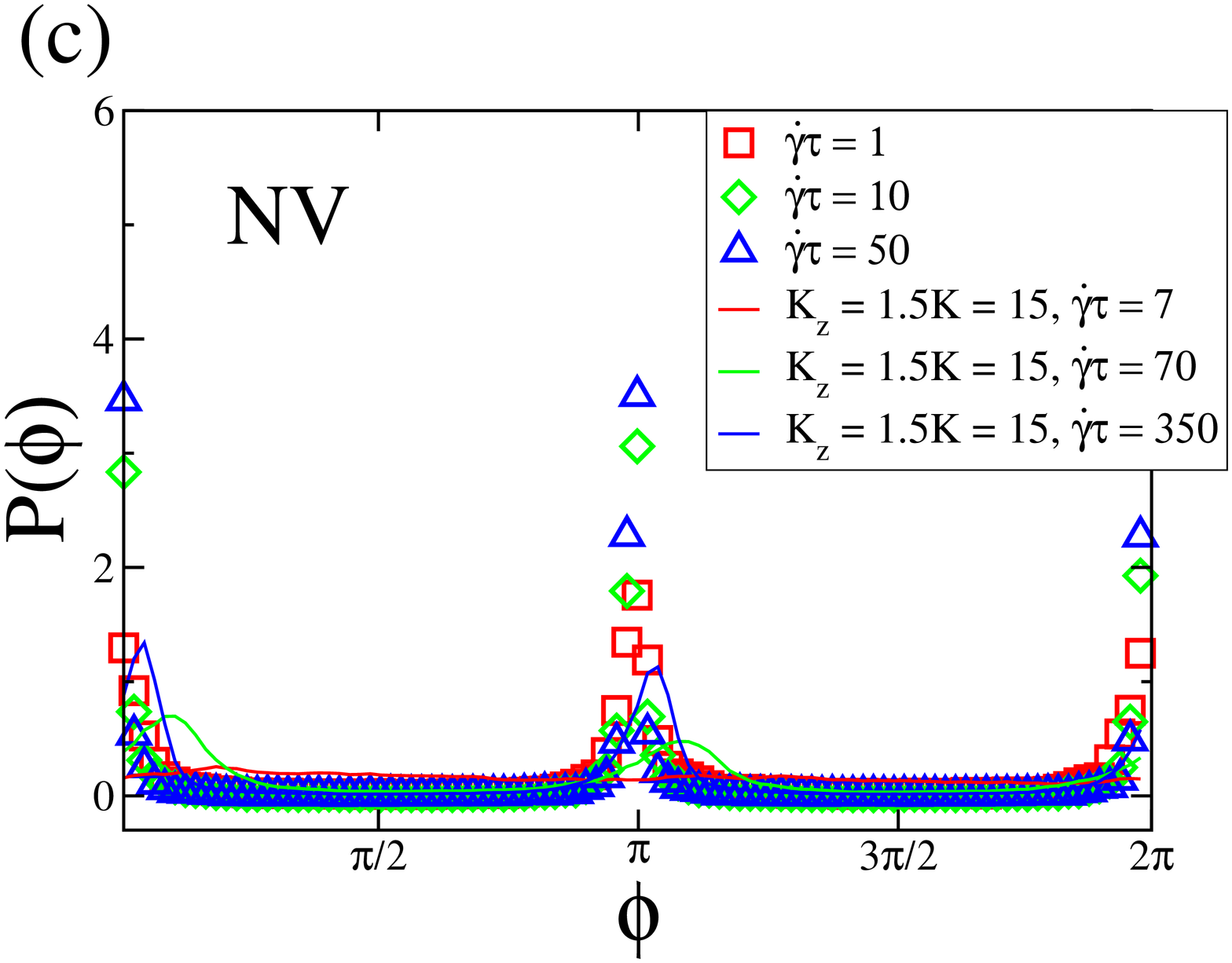}
\includegraphics[angle=270,width=0.46\columnwidth]{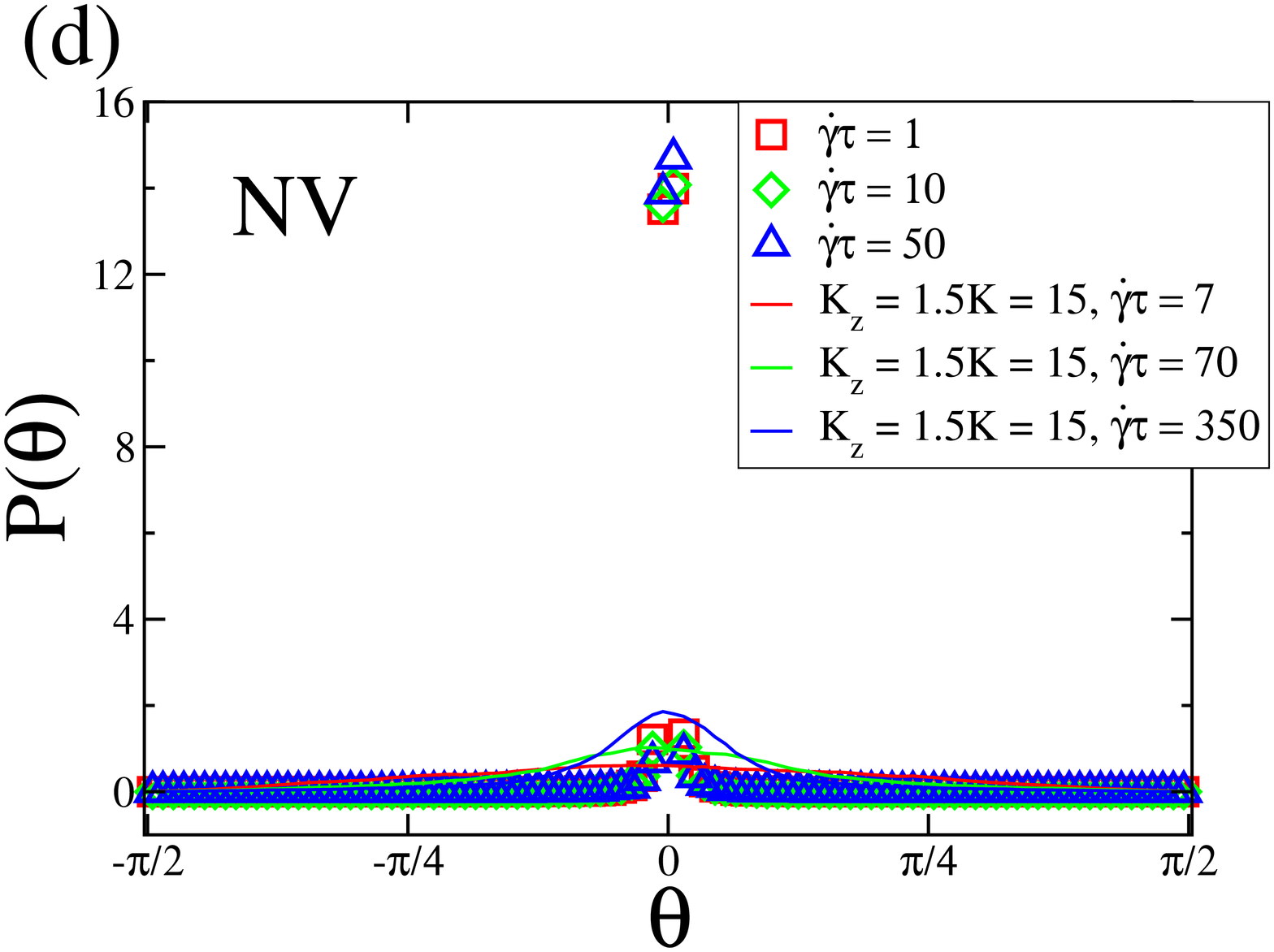}
}
\centerline{
\includegraphics[angle=270,width=0.46\columnwidth]{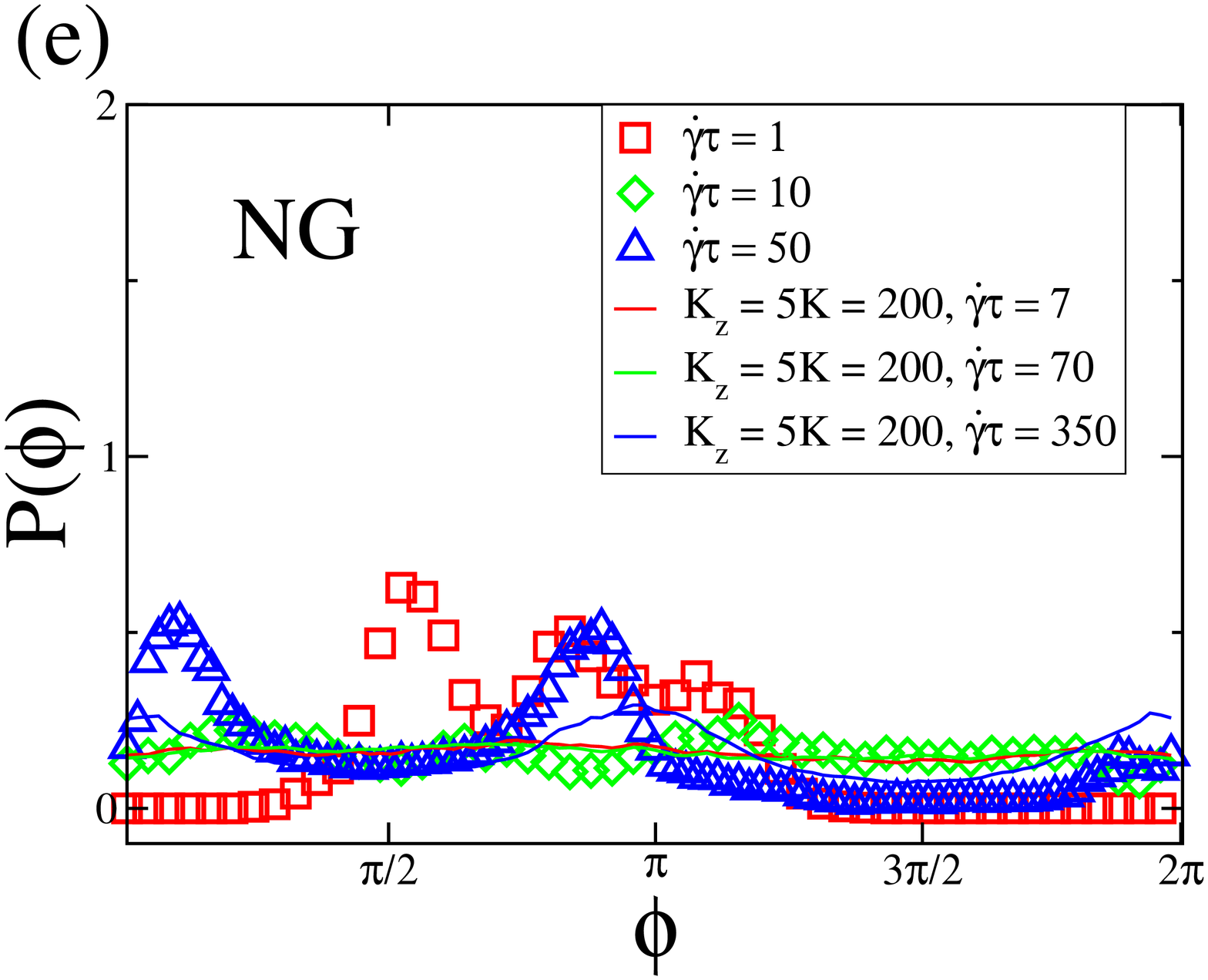}
\includegraphics[angle=270,width=0.46\columnwidth]{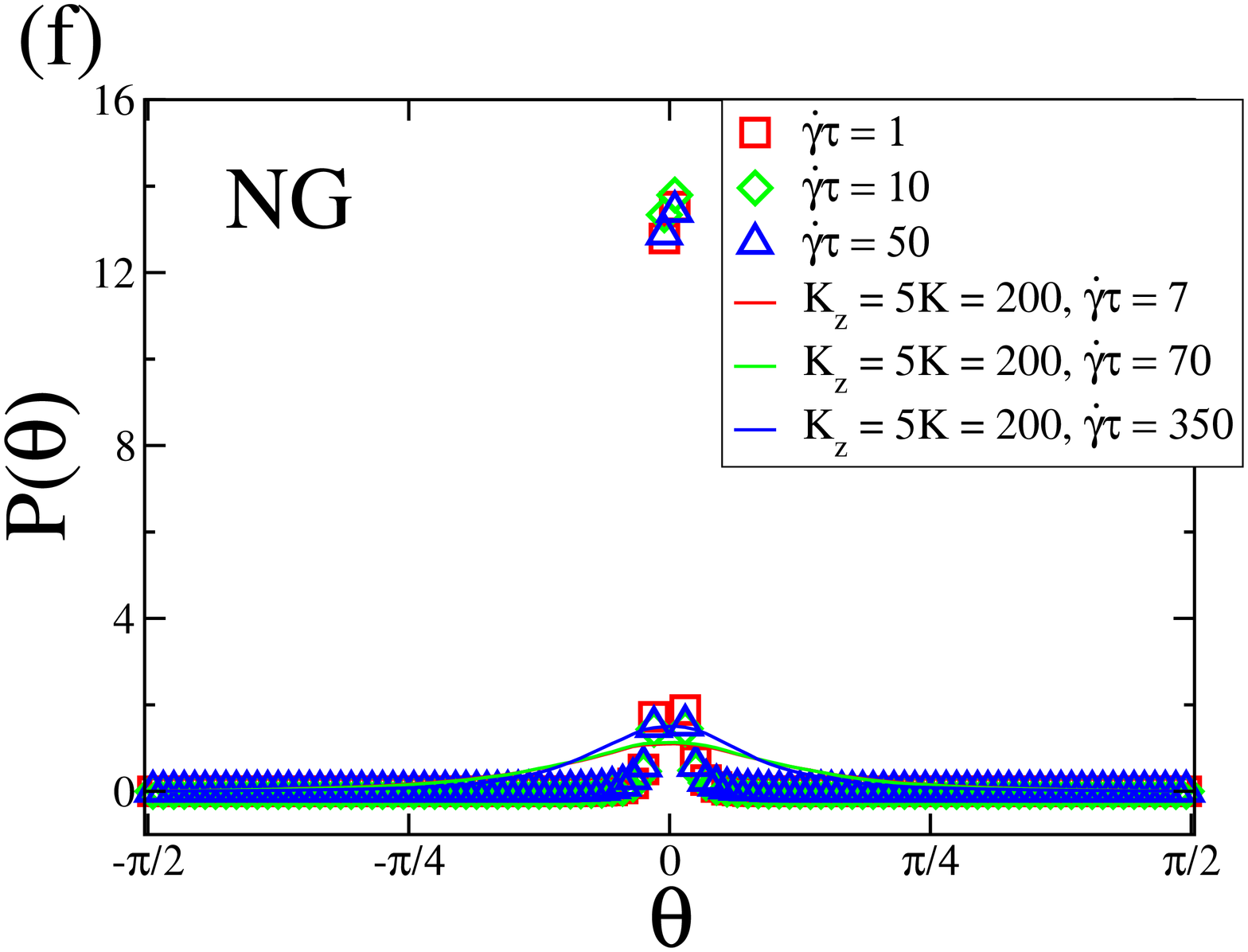}
}
\caption{The PDF of the end-to-end vector angles  (see Fig. \ref{fig:sketch}(c))
for the three cases (bulk, NV and NG). The symbols and the curves stand for the 
same parameters as those used in Fig. \ref{fig:pdfRe}.
\label{fig:angle_vs_fene}}
\end{figure}
%%%%%%%%%%%%%%%%%

%%%%%%%%%%%%%%%%%
\begin{figure}[t]
\centerline{
\includegraphics[angle=270,width=0.58\columnwidth]{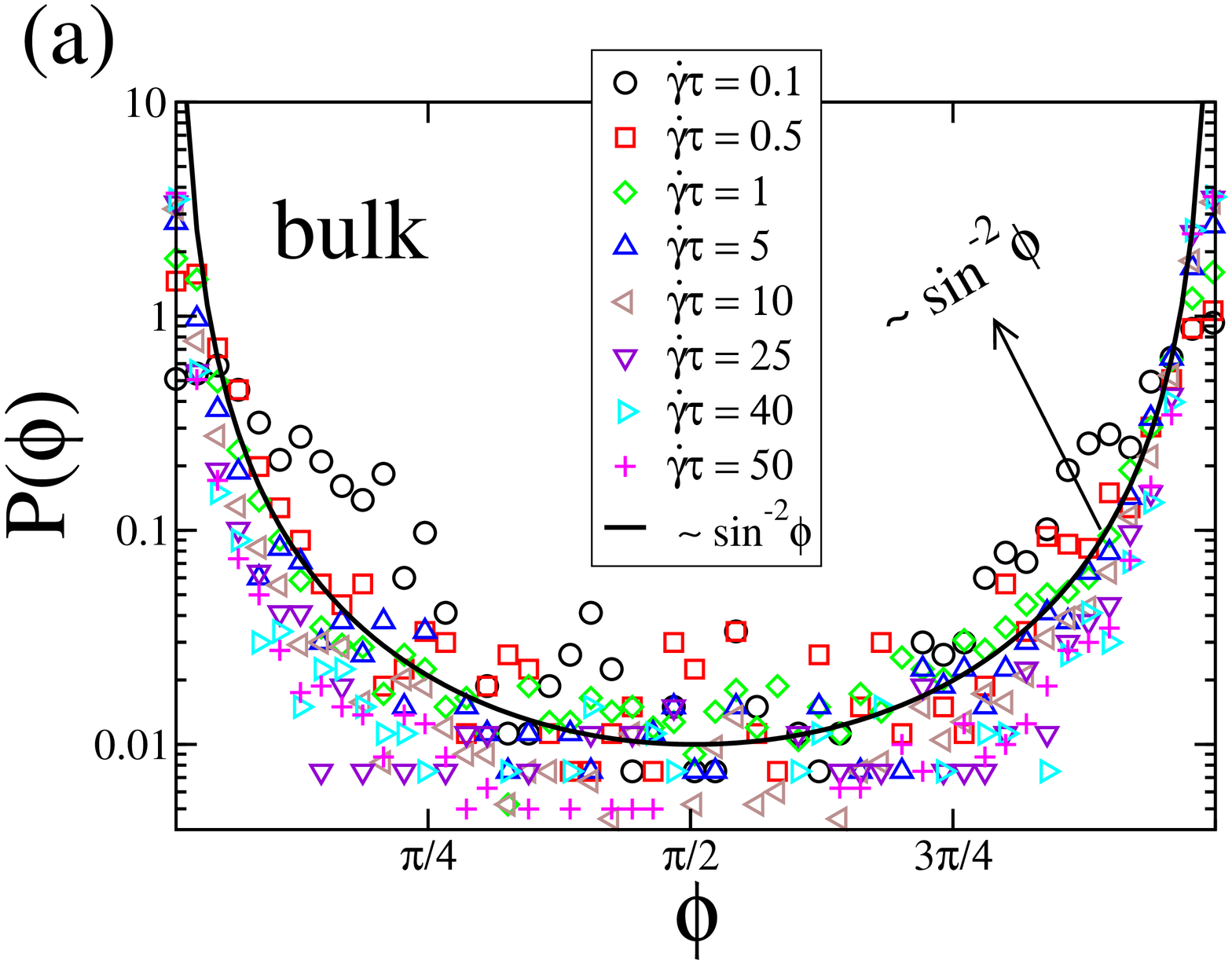}
\includegraphics[angle=270,width=0.58\columnwidth]{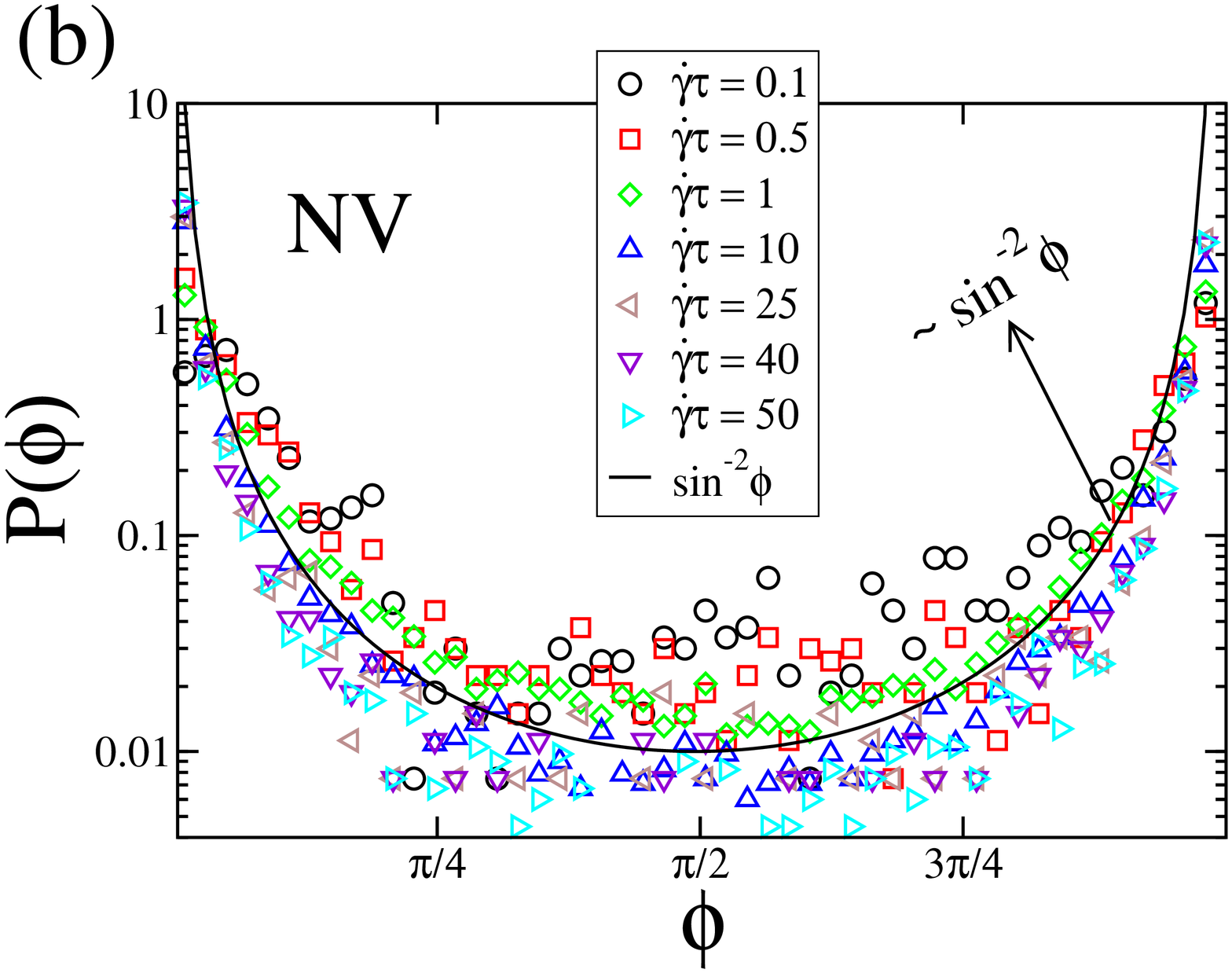}
}
%\centerline{
%\includegraphics[angle=270,width=0.89\columnwidth]{y_more_rate_P_phi_tail_fit.ps}
%\includegraphics[angle=270,width=0.5\columnwidth]{y_P_theta_tail_fit.ps}
%}
\centerline{
\includegraphics[angle=270,width=0.6\columnwidth]{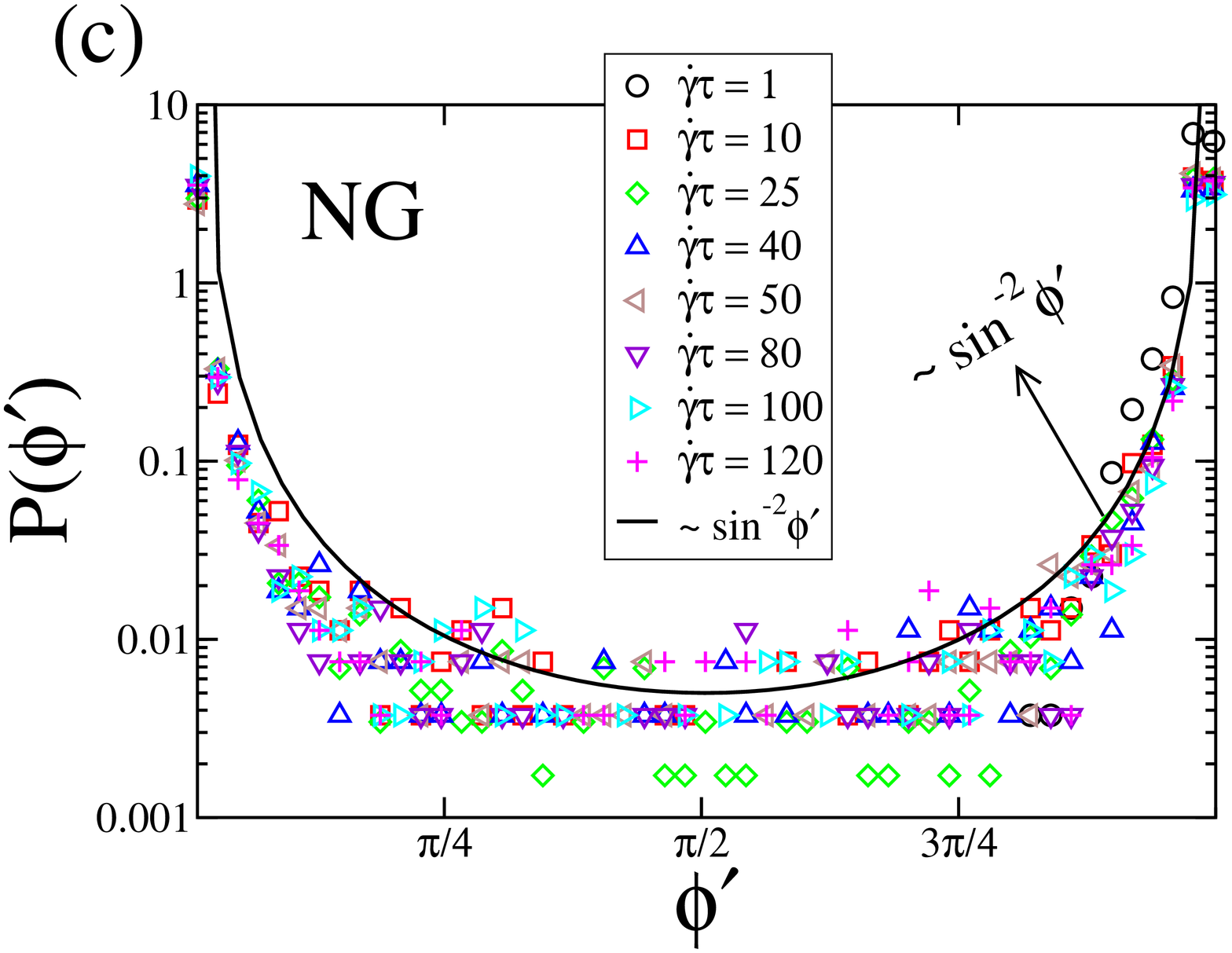}
}
\caption{The PDF of the in-shear-gradient-plane  angle 
$\phi$ ($\phi'$)  (sketched in Fig. \ref{fig:sketch}(c) and in
Fig. \ref{fig:sketch}(d)).
The solid curves are the theoretical scaling predictions ($\sim$ sin$^{-2}\phi$ or sin$^{-2}\phi'$) 
of the dumbbell model. 
\cite{Chertkov_JFM05_531,Puliafito_PD05_211} 
\label{fig:phipdf_tail}}
\end{figure}
%%%%%%%%%%%%%%%%%

%%%%%%%%%%%%%%%%%%
\begin{figure}[t]
\centerline{
\includegraphics[angle=270,width=0.58\columnwidth]{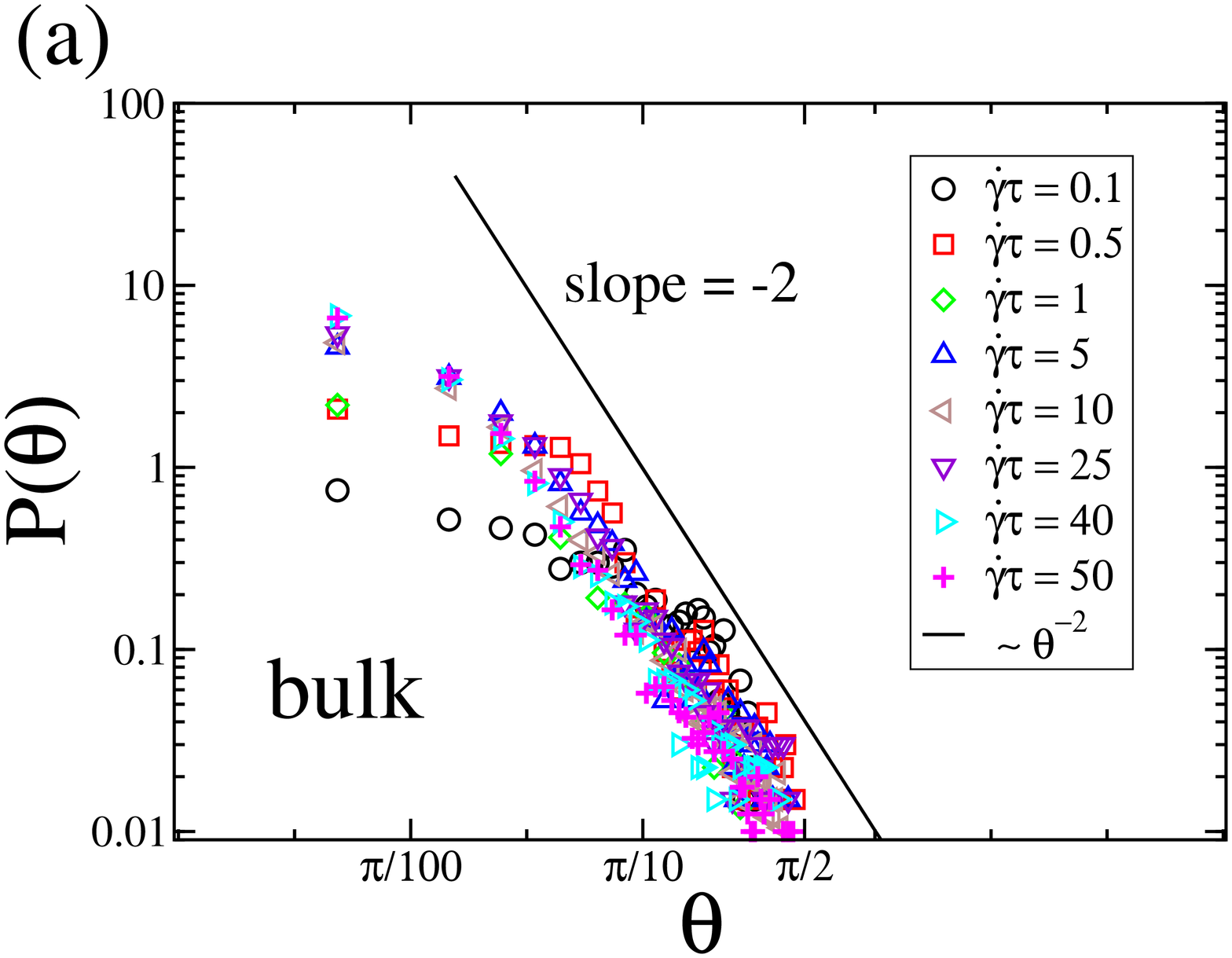}
\includegraphics[angle=270,width=0.58\columnwidth]{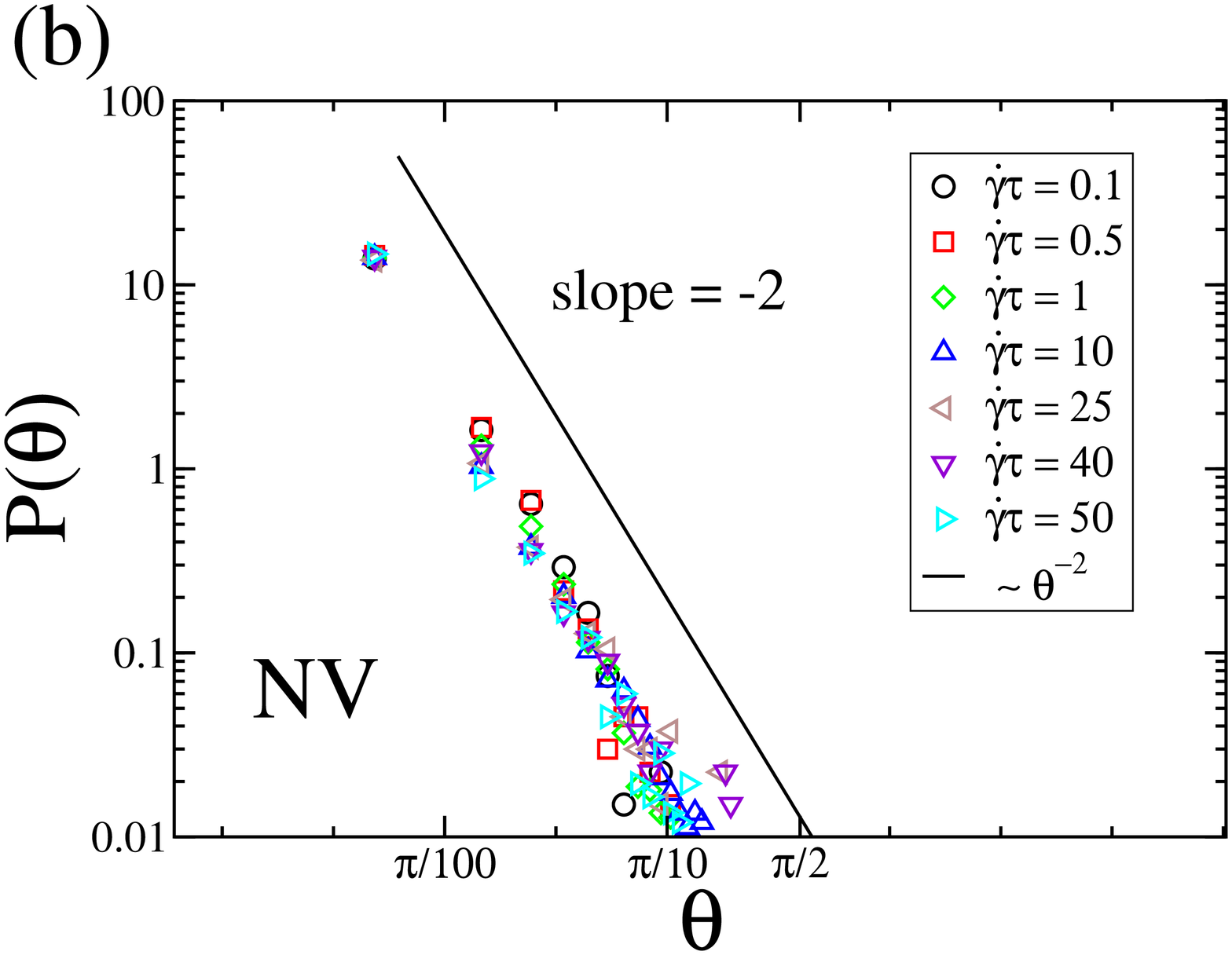}
}
%\centerline{
%\includegraphics[angle=270,width=0.5\columnwidth]{y_more_rate_P_phi_tail_fit.ps}
%\includegraphics[angle=270,width=0.89\columnwidth]{y_P_theta_tail_fit.ps}
%}
\centerline{
\includegraphics[angle=270,width=0.6\columnwidth]{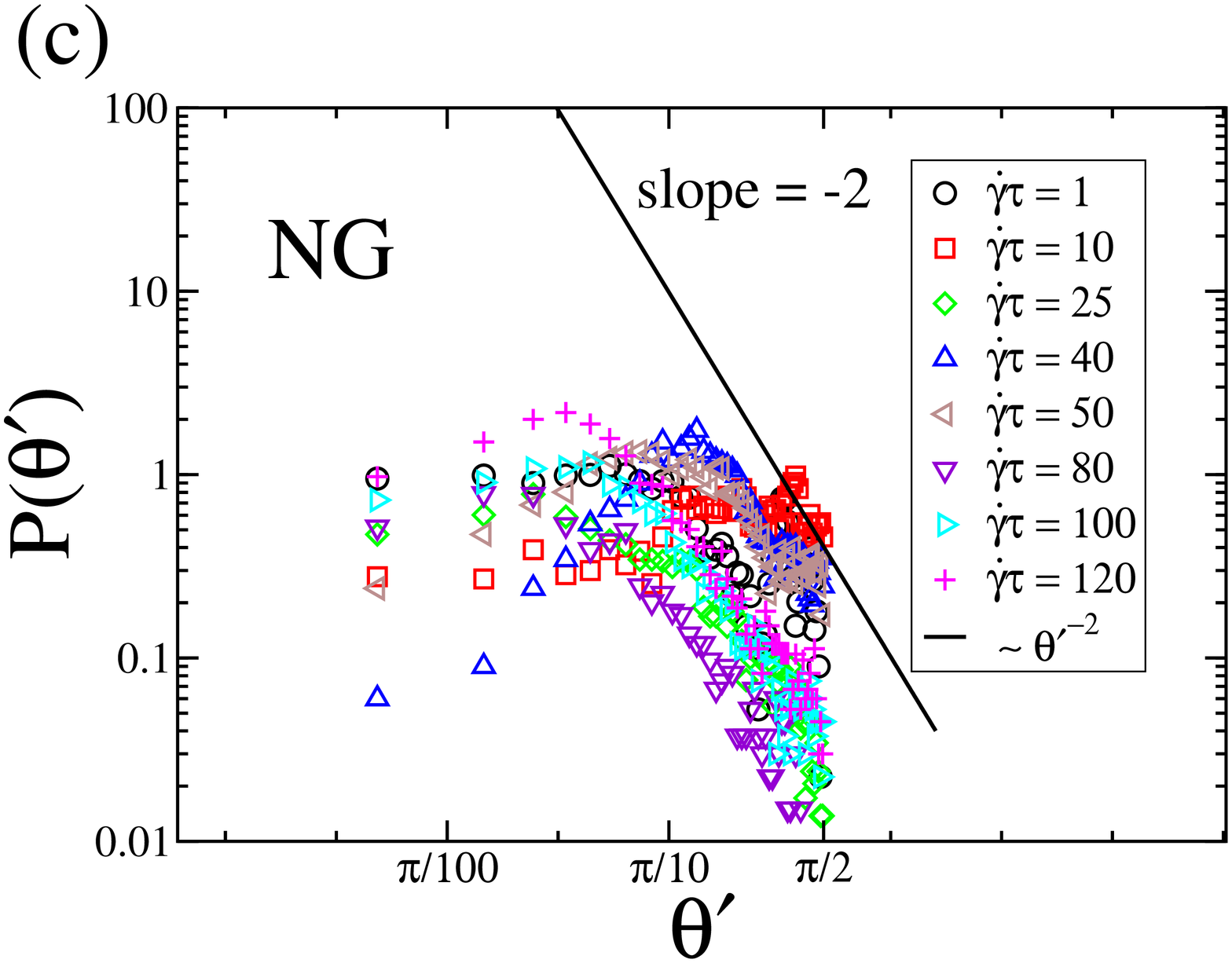}
}
\caption{The PDF of the off-shear-gradient-plane  angle 
$\theta$ ($\theta'$) (sketched in Fig. \ref{fig:sketch}(c) and in
Fig. \ref{fig:sketch}(d)).
The solid curves are the theoretical scaling predictions ($\sim$ $\theta^{-2}$ or $\theta'^{-2}$)
of the dumbbell model. \cite{Chertkov_JFM05_531,Puliafito_PD05_211}
\label{fig:thetapdf_tail}}
\end{figure}
%%%%%%%%%%%%%%%%%%

We now address the angular degrees of freedom of the end-to-end polymer chain vector, 
namely $\phi$ and $\theta$ (depicted in Fig. \ref{fig:sketch}(c)), or $\phi'$ and $\theta'$
(depicted in Fig. \ref{fig:sketch}(d)), 
with respect to the prescribed shear flow $x$-direction
as well as the shear gradient direction. 
Whereas $\phi/\phi'$ corresponds to the usual azimuthal angle, 
$\theta/\theta'$ is not the polar angle but rather 
the angle formed between $\vec R_e$ and the plane containing
the shear- and the gradient-direction, see Fig. \ref{fig:sketch}(c,d).
Thereby we have kept the same (well established) definitions used in earlier publications.
\cite{Hinch_JFM72_52, Chertkov_JFM05_531}
The results for the angular distribution functions are shown 
in Fig. \ref{fig:angle_vs_fene}.
As far as the azimuthal angle $\phi$ distribution is concerned,
BD simulation data as well as the dumbbell model 
show a strong probability peaks in the vicinity of $\phi = 0, \pi, 2\pi$ in the bulk 
(Fig. \ref{fig:angle_vs_fene}(a)) and the NV case (Fig. \ref{fig:angle_vs_fene}(c)).
Interestingly the corresponding peak heights are not identical and,  in fact,
decrease with growing angles. 
The simple underlying physical mechanism accounting for this effect is that the torque, 
(resulting from the shear gradient \cite{Schroeder_PRL_2005}),
exerted on the chain favors more the conformations with $\phi \to 0$ than those with 
$\phi \to 2\pi$. This finding of the asymmetry in $P(\phi)$, although unexplained, 
was also reported in the earlier work of Celani et al. \cite{Celani_EPL05_464}.
Moreover, the larger the shear rate $\dot \gamma \tau$, 
the narrower the density probability becomes around the flow ($x$-axis) direction,
see Fig.  \ref{fig:angle_vs_fene}(a,c). 
However, the PDFs are somewhat different for the NG case, see Fig. \ref{fig:angle_vs_fene}(e),
due to the shear gradient direction being off-plane hereby.

As far as the $\theta-$PDFs are concerned, the  shear flow
induces a concentrated distribution around $\theta=0$ for all 
shear flow geometries, see Fig. \ref{fig:angle_vs_fene}(b,d,f).
$^{[{\bf{Note}}\ 3]}$
\footnotetext[3]{
Note that the symmetry of the shear flow imposes a symmetry
of $P(\theta)$ about $\theta = 0$.}
The narrowness of the peak increases again with the shear rate, 
as expected.

%characterized by larger $\dot \gamma \tau$ narrower PDF in the bulk
%(see Fig. \ref{fig:angle_vs_fene}(b)). 
 
In a general manner, the agreement between our simulation data and the 
dumbbell model (see Fig. \ref{fig:angle_vs_fene}) is not as satisfactory 
as that found for the size distribution (compare with Fig. \ref{fig:pdfRe}). 
This is quite acceptable, given the crudeness of the dumbbell model.
Nonetheless, the essential feature of strong depletion zones in the angular PDFs, 
are qualitatively well captured by this dumbbell model.

The angular distribution functions were also recently studied analytically
by Chertkov et al.  \cite{Chertkov_JFM05_531} for a linear dumbbell in the bulk. 
It was thereby found to obey the following scaling
law \cite{Chertkov_JFM05_531} $P(\phi) \sim \sin^{-2} \phi$, 
which was experimentally confirmed by Gerashchenko and Steinberg.  
\cite{Gerashchenko_PRL_2006}
%under assumption of that
%the effect of velocity fluctuations is stronger than that related to thermal
%noise, an algebraic behavior of the angular PDF tails was well investigated
%for the polymer in the bulk state.
To establish a comparison between our simulation data and  this theoretical prediction, 
the PDFs are plotted for a wider range of the shear rate on a linear-logarithmic plane
in Fig. \ref{fig:phipdf_tail}.
A good agreement is found between the simulation and the scaling prediction.
This is a remarkable result, since a simple harmonic dumbbell model can capture
the correct scaling behavior of $P(\phi)$ of a full chain.

Concerning the $\theta$ distribution function, 
Chertkov et al.  \cite{Chertkov_JFM05_531} have also 
shown a scaling behavior for the tail of the form $P(\theta) \sim \theta^{-2}$  
within the framework of a linear dumbbell in the bulk.
Comparison in the past with experimental results \cite{Gerashchenko_PRL_2006}
demonstrated already an excellent agreement.
Our results for the $\theta$-PDF can be found in Fig. \ref{fig:thetapdf_tail}.
The scaling prediction is again in remarkable good agreement with our BD data.
Interestingly, the agreement seems to be even better for the NV case, 
where not only the tail is dictated by $\theta^{-2}$ but 
the full $\theta$-range. This suggests that the algebraic decay
$P(\theta) \sim \theta^{-2}$ is even more relevant for (nearly) 
two-dimensional systems.
%
%and in Fig. \ref{fig:thetapdf_tail}.
%
%For the NG case, the PDF tails of both, i
%$\theta'$, can only fit to the theory in even higher shear rates 
%(see Fig. \ref{fig:phipdf_tail}(c) and Fig. \ref{fig:thetapdf_tail}(c)) comparing to the
%simulation results in the bulk (see Fig. \ref{fig:phipdf_tail}(a) and 
%Fig. \ref{fig:thetapdf_tail}(a)). This difference can be explained by that,
%due to the strong adsorption, the flow with low rates can not win the dominated
%contribution to the polymer motion, such that the chain takes most of the time to
%fluctuate in the wall plane, and in the same sense, 
%the chain gets very less chance to act to the weak vertical gradient flow.     

%%%%%%%%%%%%%%%%%%%%%%%%%%%%%%%%%%%%%%%%%%%%%%%%%%%%%%%
\subsection{Bond-bond angle statistics}
\label{sec:BBangle}
%%%%%%%%%%%%%%%%%%%%%%%%%%%%%%%%%%%%%%%%%%%%%%%%%%%%%%%
%

%%%%%%%%%%%%%%%%%
\begin{figure}[t]
\centerline{
\includegraphics[angle=270,width=1.0\columnwidth]{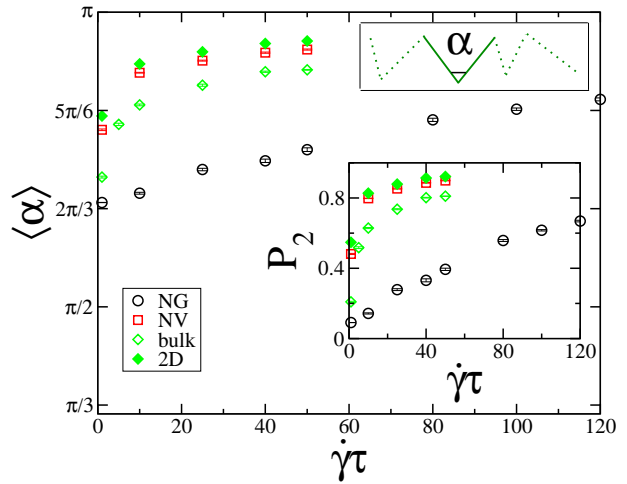}
}
%\centerline{
%\includegraphics[angle=270,width=1.0\columnwidth]{mean_cos_alpha_3cases_err.ps}
%}
\caption{The average bond-bond angle $\left<\alpha\right>$ 
plotted against the reduced shear rate $\dot \gamma \tau$. 
Inset: The order parameter $P_2(\cos \alpha)$. 
%$P_2 = \frac{3}{2}\left(\left<\cos^2(\pi - \alpha)\right> - \frac{1}{3}\right)
%=\frac{3}{2}\left(\left<\cos^2\alpha\right> - \frac{1}{3}\right)$. \cite{he_MM07_40_6721}
\label{fig:BBA}}
\end{figure}
%%%%%%%%%%%%%%%%%

%%%%%%%%%%%%%%%%%
\begin{figure}[t]
\centerline{
\includegraphics[angle=270,width=0.58\columnwidth]{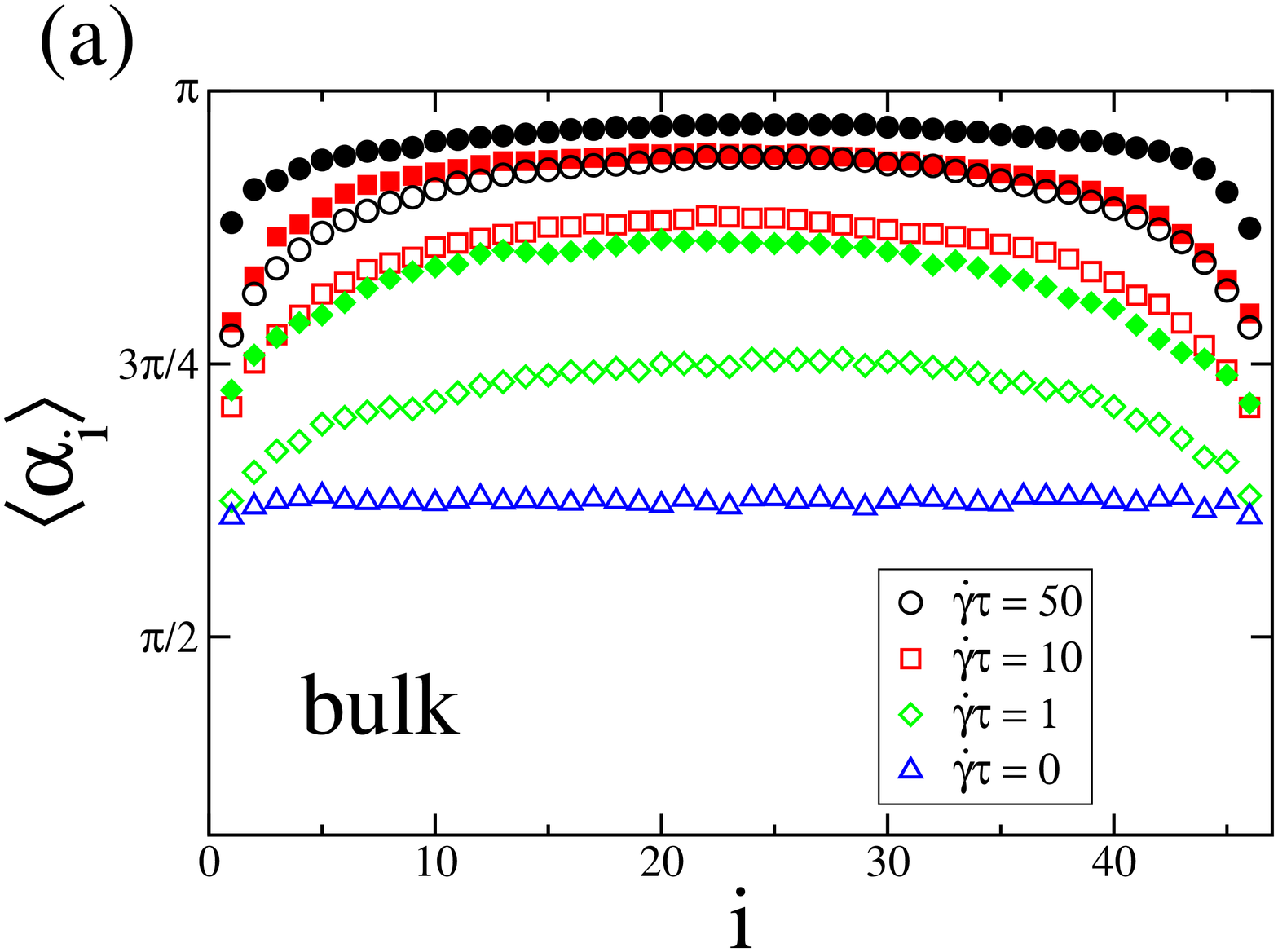}
\includegraphics[angle=270,width=0.58\columnwidth]{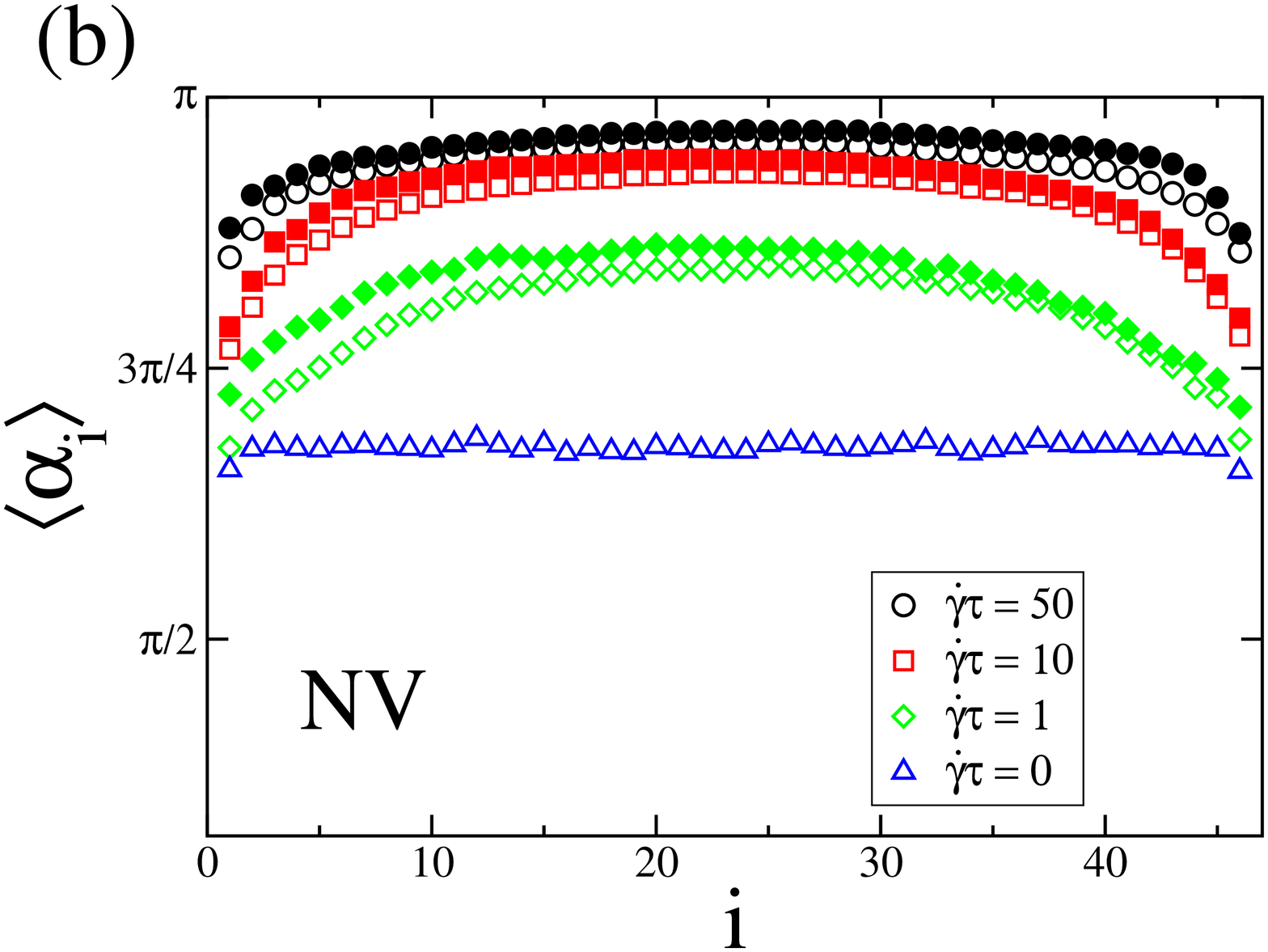}
}
%\centerline{
%\includegraphics[angle=270,width=0.89\columnwidth]{y_alpha_IDmono.ps}
%}
\centerline{
\includegraphics[angle=270,width=0.6\columnwidth]{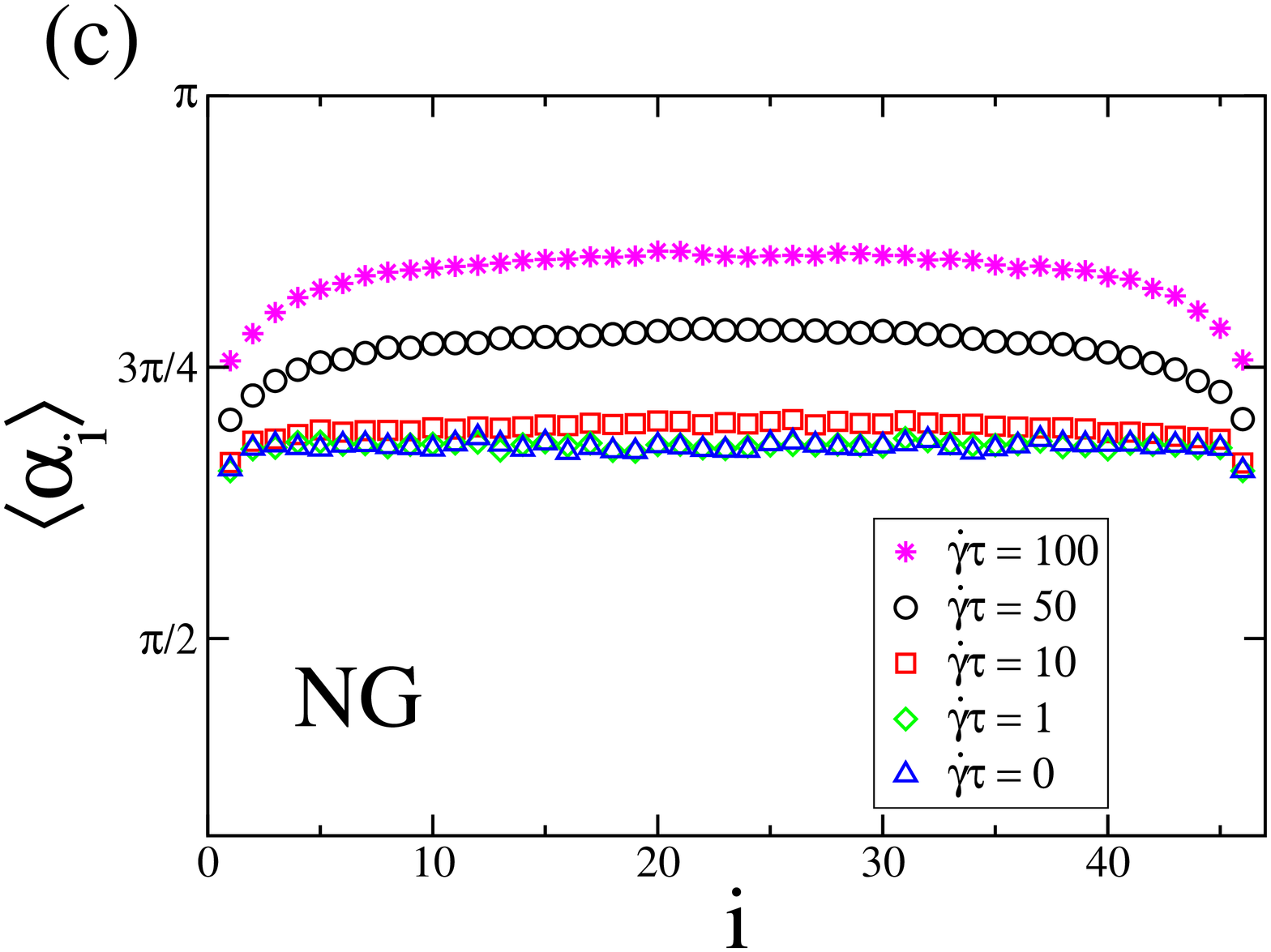}
}
\caption{The mean local bond-bond angle $\left<\alpha_i\right>$ 
of monomer ($i$). The solid symbols in (a) and (b) represent
the purely 2D case.
\label{fig:BBA_vs_ID}}
\end{figure}
%%%%%%%%%%%%%%%%%

To characterize the intra chain property under shear flow, 
we have considered the bond-bond angles (BBAs). 
This is a complementary approach to investigate the chain stretching and coiling features.
The results are depicted in Fig. \ref{fig:BBA}. 
In a first low shear rate window ($\dot \gamma \tau \lesssim 20$), 
there is a rather strong shear rate dependency for the NV, bulk, and 2D cases.
Typically, this describes the relevance of coiled conformations.
At higher shear rates, this dependency becomes much weaker
and a plateau-like regime (about $\pi$) is observed.
This feature, is the signature of strongly stretched polymer chains.
As far as the NG case is concerned, the shear rate dependency is much weaker
as expected.
This result is fully consistent with the positive skewness reported in 
Fig. \ref{fig:momentsRe}(c) for a wide range of the shear rate.
A complementary observable provided by the following order
parameter \cite{he_MM07_40_6721} $P_2$ is also shown as an inset in Fig. \ref{fig:BBA}.

\begin{equation}\label{eq:P2}
P_2(\cos \alpha) = \frac{3}{2}\left(\left<\cos^2(\pi - \alpha)\right> - \frac{1}{3}\right)
 = \frac{3}{2}\left(\left<\cos^2\alpha\right> - \frac{1}{3}\right) \;.
\end{equation}   
%
%In the inset of Fig.
%\ref{fig:BBA}, this quantity is plotted against the shear rate for the different shear environments.
%On the one hand, with increasing the shear rate in the inset, 
%the bond rotation becomes further 
%restricted to orient to the flow direction due to the larger shear strength
%winning in the competition with fluctuations, resulting of the observed increase
%of $P_2$. On the other hand, in the 2D-like NV case, the orientation freedom of the bond
%is the most restricted by the flow, followed by less restricted in the bulk,
%and the least confined in the NG case. 
%Due to the strong adsorption, the chain is almost confined
%in the wall plane, such that, in the NG case (vertical shear gradient), the 
%chain couldn't feel much shear strength, which creates the smallest value of 
%$P_2$ in this case; in the NV (in-wall-plane gradient), since the monomers
%fluctuate in the shear gradient direction much more than those in the bulk.
%the same shear rate imposes more strength on the monomers in the NV,
%which results of stronger chain stretching to create the largest $P_2$ in this case.

We now would like to address the local bond-bond angles, 
which physically characterizes the degree of local chain stretching.
The mean BBA $\left<\alpha_i\right>$
is plotted {\it{vs}.} the monomer position index ($i$) in Fig. \ref{fig:BBA_vs_ID}. 
As expected, a symmetry around the chain center is revealed, see Fig. \ref{fig:BBA_vs_ID}.
The two chain ends are always looser and the {\it local} stretching,  
is smoothly increasing upon approaching the center of the chain.
Besides, the local chain stretching increases with growing shear rate, 
see Fig. \ref{fig:BBA_vs_ID}.

%%%%%%%%%%%%%%%%%%%%%%
\subsection{Tumbling}
\label{sec:frequency}
%%%%%%%%%%%%%%%%%%%%%%

%%%%%%%%%%%%%%%%%%
\begin{figure}[t]
\centerline{
\includegraphics[angle=270,width=1.0\columnwidth]{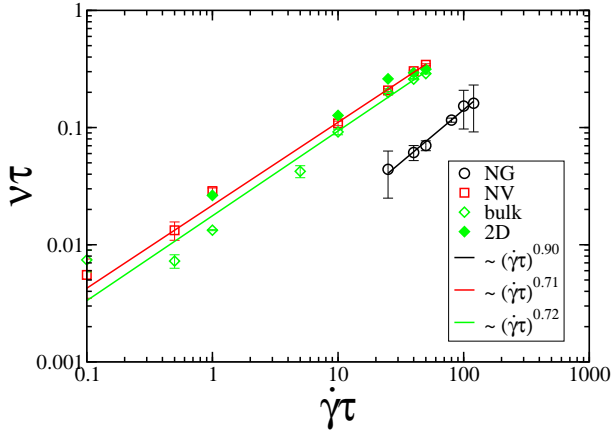}
}
\caption{Reduced tumbling frequency $\nu \tau$ as a function of the reduced shear rate $\dot \gamma \tau$.
The lines correspond to the best fit.
\label{fig:frequency}}
\end{figure}
%%%%%%%%%%%%%%%%%%

% \begin{figure}[t]
% \centerline{
% \includegraphics[angle=270,width=1.0\columnwidth]{fluctuation_in_grad_dir_err.ps}
% }
% \caption{The mean monomer fluctuations in the shear gradient direction ($y-$direction for 
% the bulk and NV; $z-$direction for the NG). The symbols have the same meaning as those
% used in Fig. \ref{fig:frequency}.
% \label{fig:fluc_grad}}
% \end{figure}

The tumbling process of the chain under shear flow is the mechanism by which
energy is transferring, or in other words, the polymer cyclically absorbs (stretching)
and releases (collapsing) energy from and to the surrounding fluid. 
This particular polymer motion in shear flow has been thoroughly studied in the past. 
\cite{Teixeira_MM05_38,Schroeder_PRL_2005,Rafael_PRL_2006}
%The tumbling frequency of the macromolecule under such flow has been obtained either by counting
%clear, end-over-end tumbling events, together with scaling argument analysis
%\cite{Teixeira_MM05_38}, or by plotting the power spectral density (PSD) 
%\cite{Press_CUP_97} of
%time-fluctuation polymer configurations for free chains 
%\cite{Schroeder_PRL_2005}, or by calculating the cross power spectral density
%(CPSD) \cite{Carter_IEEE79} associated with the size extension in flow and
%shear gradient direction for a grafted polymer \cite{Rafael_PRL_2006}. 
All these studies address  
the bulk case \cite{Teixeira_MM05_38,Schroeder_PRL_2005} or grafted polymer chains, 
\cite{Rafael_PRL_2006} but without considering the problem of adsorption.
In this work, the frequency of this cyclic motion is calculated for
the three geometries (bulk, NV and NG) via
the Fourier transform of the time-dependent radius of gyration,
which is
%
%%%%%%%%%%%%%%%%
\begin{equation}
\label{gyration}
R_g(t) = \sqrt{\frac{\sum^{N-1}_{i = 0}({\bf{r}}_i(t) - {\bf{r}}_{c.m.}(t))^2}{N}}\;,
\end{equation} 
%%%%%%%%%%%%%%%%
%
where ${\bf{r}}_{c.m.}(t) = \sum_{i = 0}^{N -1}{\bf{r}}_i(t)/N$ 
is the chain center of mass at time $t$. 
The first main peak position of the resulting Fourier transform of $R_g(t)$ (not shown here)
is chosen as  the tumbling frequency. 

Figure \ref{fig:frequency} displays the reduced tumbling frequency 
($\nu \tau$) as a function of the reduced shear rate $\dot \gamma \tau$. 
The purely 2D case is also shown as a reference.
Thereby, scaling behaviors of the form 
$$
\nu \tau \sim (\dot \gamma \tau)^{\lambda} 
$$ 
are observed , with $\lambda = 0.90, 0.71, 0.72$ for NG, NV, and bulk, respectively. 
Hence, the bulk and NV cases, exhibit (within statistical uncertainties) an
identical scaling behavior for the tumbling frequency.
The value found for the exponent 
$\lambda \approx 0.7$ is in very good agreement with the one
reported by Schroeder et al. \cite{Schroeder_PRL_2005} ($\lambda = 0.67$),
based on the angular power spectral frequency, for Brownian rods in the bulk.
As expected the NG case exhibits a different scaling behavior (here $\lambda=0.9$)
due to the chain adsorption that weakens the effect of the velocity gradient in the 
normal direction.
%This figure indicates that, in the range of the plotted shear rates, the tumbling 
%frequency increases monotonically with shear strength, and scales sublinearly
%with flow rate.

To understand the physical origin of the observed scaling behaviour of the tumbling
frequency in Fig. \ref{fig:frequency}, 
we are going to employ the arguments advocated by Teixeira et al. 
\cite{Teixeira_MM05_38}. 
By decomposing the tumbling motion into typical time periods:
(i) a  stretching phase ($t_{stretch}$), (ii) an alignment along the shear direction 
($t_{align}$), (iii) a flip motion ($t_{flip}$), 
and (iv) a collapse event ($t_{collapse}$),
we can write an expression for the tumbling frequency 
\cite{Schroeder_PRL_2005,Teixeira_MM05_38}  $\nu$
%
%%%%%%%%%%%%%%%%
\begin{equation}
\label{eq:frequency}
\nu \tau \propto \frac{\tau}{t_{stretch} + t_{align} + t_{flip} + t_{collapse}}\;. 
\end{equation}  
%%%%%%%%%%%%%%%%
% 
%able to be explained with the help of developing a physical description by
%deriving a frequency equation from simple advection and diffusion mechanism, where
%the characteristic time scales of the tumbling process is dissected into four 
%distinct phases: stretching, aligning, flipping and collapsing
%\cite{Schroeder_PRL_2005, Teixeira_MM05_38}. This scaling study is based on 
%analysing the simplest polymer model, the dumbbell model and the equation
%of the frequency is given below \cite{Schroeder_PRL_2005, Teixeira_MM05_38}:
%
The three processes of polymer stretching, aligning and collapsing being advection-driven, 
Teixeira {\it{et al.}} \cite{Teixeira_MM05_38} have shown that 
$t_{stretch}$, $t_{align}$, and $t_{collapse}$ scale like
$\frac{\left<X\right>}{(\dot \gamma \tau)\left<\delta_2\right>}$, where
$\left<X\right>$ is the average chain extension in the $x$-flow-direction
and $\left<\delta_2\right>$ is the chain extension in the shear gradient direction.
%
%Here, $\left<\delta_2\right>$ cooresponds to $\left<Y\right>$ is  
%in the $y-$direction for the bulk and NV, $\left<Z\right>$ 
%in the $z-$direction for the NG in our simulations. 
On the other hand, the flipping process
is a diffusive motion such that $t_{flip} \sim \frac{\left<\delta_2\right>^2}
{D(\delta_2)} \sim \left<\delta_2\right>^{8/3}$, \cite{Teixeira_MM05_38}
where $D(\delta_2)$ is the diffusivity in the shear gradient direction 
and found to be $D(\delta_2) \sim \delta_2^{-2/3}$ from BD simulations
by Doyle {\it{et al.}}. \cite{Doyle_JFM97_334_251}
Following the same procedure with our simulation data, 
we were then able to extract the  time duration  scalings 
which are gathered  in table \ref{tab:table1}.

%
%%%%%%%%%%%%%%%
\begin{table}[h]
\centering
\caption{Characteristic times involved in the tumbling motion discussed in the text.} 
\label{tab:table1}
\begin{tabular}{l | c | c | c | c}
\hline
\hline
     & $t_{stretch}$ & $t_{align}$ & $t_{collapse}$ & $t_{flip}$ \\
\hline
bulk &$\sim (\dot \gamma \tau)^{-0.62}$&$\sim (\dot \gamma \tau)^{-0.62}$&$\sim (\dot \gamma \tau)^{-0.62}$&$\sim (\dot \gamma \tau)^{-0.67}$\\
NV  &$\sim (\dot \gamma \tau)^{-0.64}$&$\sim (\dot \gamma \tau)^{-0.64}$&$\sim (\dot \gamma \tau)^{-0.64}$&$\sim (\dot \gamma \tau)^{-0.61}$\\
NG  &$\sim (\dot \gamma \tau)^{-0.87}$&$\sim (\dot \gamma \tau)^{-0.87}$&$\sim (\dot \gamma \tau)^{-0.87}$&$\sim (\dot \gamma \tau)^{0.40}$\\
\hline
\hline
\end{tabular}
\end{table}
%%%%%%%%%%%%%%%

These results in Table \ref{tab:table1} are
%the veracity of 
%the simple scaling arguments used by Teixeira \cite{Schroeder_PRL_2005,Teixeira_MM05_38},
%given the 
in good agreement, for the NV case and the bulk, with the observed exponent ($\lambda=0.7$, see Fig. \ref{fig:frequency}).
%Using the simulation data into the predicted time scaling with the shear rate,
%the data in table \ref{tab:table1} demonstrate that the simulation results
%are in good agreement with the scaling arguments of the tumbling frequency
%for the bulk and NV cases, 
Even for the NG case, the scaling  predictions are validated for the convective processes
(stretching, alignment, collapsing).
Nonetheless, it fails for $t_{flip}$ in the NG case because
the strong adsorption weakens 
the diffusivity in the shear gradient ($z-$) direction $D(\delta_2)$.
%is no longer the one obtained in the bulk state \cite{Doyle_JFM97_334_251}.   

% The periodic time difference in the three cases is that, in Fig.
% \ref{fig:frequency}: the cyclic motion in the NV case goes fastest (open squares), 
% slightly slower in the bulk (open diamonds) and the slowest in the NG 
% (open circles). To understand the different absolute magnitudes of the 
% frequency in the three situations, the average monomer fluctuations 
% in the shear gradient direction is computed and plotted in 
% Fig. \ref{fig:fluc_grad} against the shear rate. The difference of the 
% tumbling motion in the bulk and the NV model takes place in the process
% of the flipping phase, where the diffusivity in the shear gradient direction
% is nearly the same, the monomer fluctuations in this same direction  
% (see the squares in Fig. \ref{fig:fluc_grad}) are slightly larger in the NV
% than those in the bulk
% (see the diamonds in Fig. \ref{fig:fluc_grad}), such that the frequencies
% are slightly smaller in the later case (see the open squares and diamonds in Fig.
% \ref{fig:frequency}). Although the tumbling motion is also observed in
% the large shear rate in the NG case, the tumbling time
% (see the circles in Fig. \ref{fig:frequency}) is getting 
% much longer since the monomer mobility in the gradient direction
% is significantly reduced (see the inset of Fig. \ref{fig:fluc_grad})
% due to the strong adsorption imposing in the same direction.      

\section{Conclusion}\label{sec:conclusion}
Using Brownian dynamics simulations, the statistical properties of adsorbed polymers
under a linear shear flow have been studied for the two cases: in-adsorption-plane 
shear gradient (or equivalently the shear vorticity normal to the adsorbing wall), 
and the shear gradient perpendicular to the wall. 
The three-dimensional bulk and  the purely two-dimensional cases under linear shear flow  have
been considered as well as reference cases.

We have compared the behaviour of the full polymer chain to 
that of a simple dumbbell model.
A good agreement is reached for the PDF of the chain
end-to-end distance. Besides,    
the theoretically predicted scaling behaviour 
\cite{Chertkov_JFM05_531,Puliafito_PD05_211} of the angular PDF tails
is confirmed  also in the case where the shear gradient is parallel to the walls (NV case).
In the opposite case of perpendicular shear gradient (NG case), agreement is only found at high shear rates.
It is remarkable that a simple dumbbell model
can capture the correct scaling behaviour of the angular PDFs' tails
of a full monomer-resolved chain. 

The chain coiling and stretching degree was  
characterized by the inner bond-bond angles. For the NV, bulk and 2D cases,
the shear rate dependency is found to be rather strong in the first low shear rate regime,
but is much weaker at higher shear rates, and a plateau-like behavior is approached once the
shear rate is high enough.
For the NG case, this dependency is much weaker for a wide range of 
shear rates.

The tumbling frequency in the bulk scales sublinearly with shear flow rate, 
which is in agreement with the early studies. \cite{Teixeira_MM05_38, Schroeder_PRL_2005} 
We found this sublinear scaling also in the adsorbed cases (NV, NG).
%(Fig. \ref{fig:frequency}). 
The theoretical scaling analysis 
based on the dumbbell model, \cite{Teixeira_MM05_38} 
was confirmed for the NV case but not for the NG case.
   
As an outlook, understanding the stretching and coiling mechanisms of polymer in flow leads
 to an improved control over chain configurations
in various circumstances of confinement,
e.g.\ in microfluidic devices. \cite{Pfohl} Stretched polymer chains
may provide valuable candidates for nanowires in microcircuits. \cite{Knobloch}
Therefore a minimalistic model, as the simple FENE dumbbell model, is
useful in order to map the behaviour of a complex system and to predict the
qualitative trends in various flow situations.

For future studies, many polymer chains, i.e. a concentrated polymer solution,
may exhibit complex dynamical behaviour due to significant slowing down
caused by  mutual chain entanglements. \cite{Varnik_JCP_02_117} It is further
interesting to mix nanoparticles and polymers 
\cite{Dzubiella_JPCM_01_13} and expose them to different flow
fields and study their collective dynamical behaviour. \cite{Usta_JCP_09_130}
Finally, the influence of hydrodynamic interactions need more consideration,
following the recent simulation method proposed in ref \cite{Hoda_JCP07_127_234902}.

\vspace{0.5cm}
{\it{Acknowledgments}} We thank A. Kiriy and M. Stamm for helpful discussions.
This work was supported by the DFG (LO 418/12-1 and SFB TR6).

%\bibliographystyle{prsty}
%\bibliography{shear}

\end{document}